\documentclass{aastex63}
\usepackage[figuresright]{rotating}

\shorttitle{Photometric investigation for Miras}
\shortauthors{Ou \& Ngeow}

\graphicspath{{./}{figures/}}

\begin{document}

\title{Difference of photometric properties between regular and non-regular Miras in the Magellanic Clouds}

\correspondingauthor{Jia-Yu Ou}
\email{m1039004@gm.astro.ncu.edu.tw}

\author[0000-0002-6928-2240]{Jia-Yu Ou}
\affil{Graduate Institute of Astronomy, National Central University, 300 Jhongda Road, 32001 Jhongli, Taiwan}
\author[0000-0001-8771-7554]{Chow-Choong Ngeow}
\affil{Graduate Institute of Astronomy, National Central University, 300 Jhongda Road, 32001 Jhongli, Taiwan}

\begin{abstract}

Mira variables are asymptotic giant branch pulsating stars with long pulsation periods and large amplitudes in optical bands. By applying the random forest algorithm to the I-band light curves for the Miras in the Magellanic Clouds, we have classified these Miras into regular Miras and non-regular Miras. Wherein non-regular Miras exhibit a long-term variation in addition to their primary pulsation periods. Our results confirm that the period-luminosity relation for maximum light has a smaller dispersion, but only occurs on the regular oxygen-rich (O-rich) Miras, which we recommend to be applied in future distance scale work. We have also collected multi-band photometry for these Miras to perform a spectral-energy-distribution (SED) fitting with stellar and dust components, showing that a significant fraction of dust is present around the non-regular Miras. According to our results, we believe that the periodic long-term variations seen in the non-regular Miras might be due to the presence of dust. 

\end{abstract}


\section{Introduction} \label{sec:intro}

Mira variables (hereafter Mira) are low and intermediate-mass pulsating stars on the asymptotic giant branch (AGB) that express large periodic variation in the optical and near-infrared (NIR) bands \citep{2012Ap&SS.341..123W}, with pulsation periods spanning from $\sim100$ to $\sim1500$~days and amplitude variations of $\Delta I > 0.8$ mag and $\Delta V > 2.5$ mag. They are also very cool red giants with effective temperatures around 3000 K and their radii in a few hundred Solar radii \citep{2021MNRAS.500.1575T}. Miras can be divided into O-rich (oxygen-rich) and C-rich (carbon-rich) \citep[for example, see][]{2001A&A...377..945C,2010ApJ...723.1195R} similar to other AGB stars.

At the beginning of the twentieth century, the correlation between the pulsation periods and apparent magnitudes existed for the classical Cepheids \citep{1912HarCi.173....1L} known as the period-luminosity (PL) relation or Leavitt Law and can be used as distance indicators. The first PL relation for Miras was found using NIR observations, as presented in \citet{1981Natur.291..303G}. Until now, several papers have also attempted to devive the PL relations for Miras can be found in the literature \citep[for example, see][]{1984MNRAS.211P..51F,1989MNRAS.241..375F,2008MNRAS.386..313W,2017AJ.154..149Y,2019ApJ...884...20B}. Furthermore, long-period variables (LPV), including Miras in the Large Magellanic Cloud, were found to exhibit several sequences of the PL relation \citep{wood1999,2007AcA....57..201S} in the optical and NIR bands, at which Miras occupied the sequence C on these PL relations. Some Miras and LPV also exhibit a significant long-term trend extending several thousand days without apparent periodic variations  \citep{1997MNRAS.288..512W}. The amplitude of these long-term trends was smaller at longer wavelengths  \citep{2003MNRAS.342...86W}.

The goal of our work is to investigate the light curves of Miras in the Large and Small Magellanic Clouds (LMC and SMC, respectively), collected from the third phase of the Optical Gravitational Lensing Experiment  \citep[OGLE-III;][]{2008AcA....58...69U} and supplemented with photometric data available from SIMBAD archive. We first classified the LMC and SMC Miras into the regular Miras and non-regular Miras in Section 2. The light curves of regular Miras can be represented as a simple sinusoidal function without additional variations. In contrast, light curves of the non-regular Miras are super-positioned of a sinusoidal function and a long-term variation. In this study, we assume the long-term variation is periodic. The main results of our analysis will be presented in Section 3 and 4 for the regular and non-regular Miras, respectively. We then performed a multi-band analysis on these Miras in Section 5, followed by conclusions given in Section 6.


\section{Light Curves Classification}

\begin{figure}
    \centering
    \includegraphics[width=1\columnwidth]{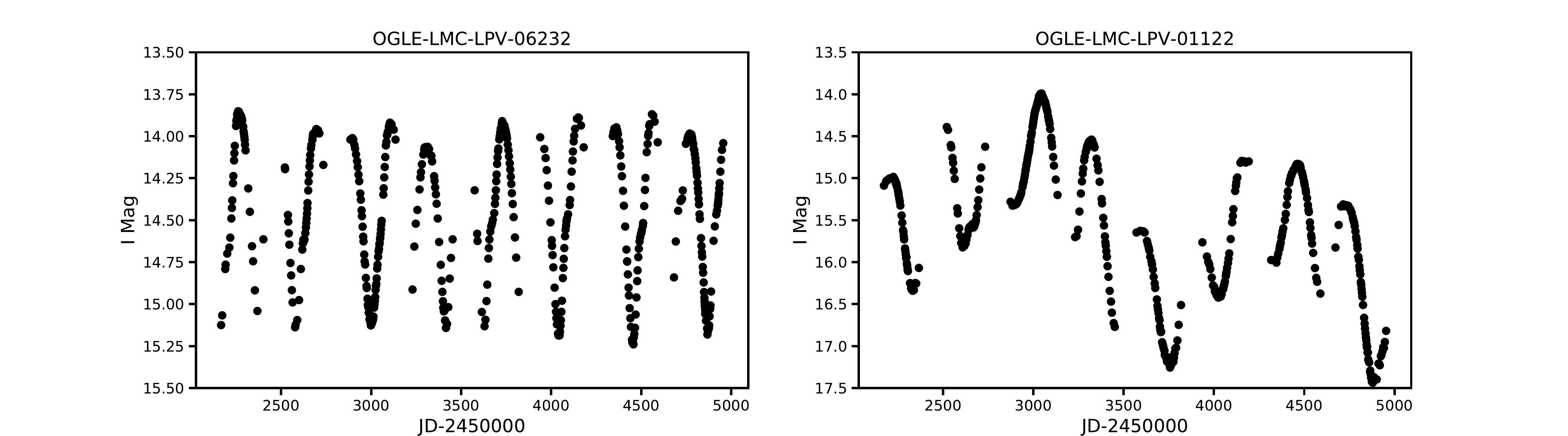}
    \caption{Examples of a regular and a non-regular Miras light curves are shown in the left and right panel, respectively.}
    \label{LC_example}
\end{figure}

\begin{figure}
    \centering
    \includegraphics[width=1\columnwidth]{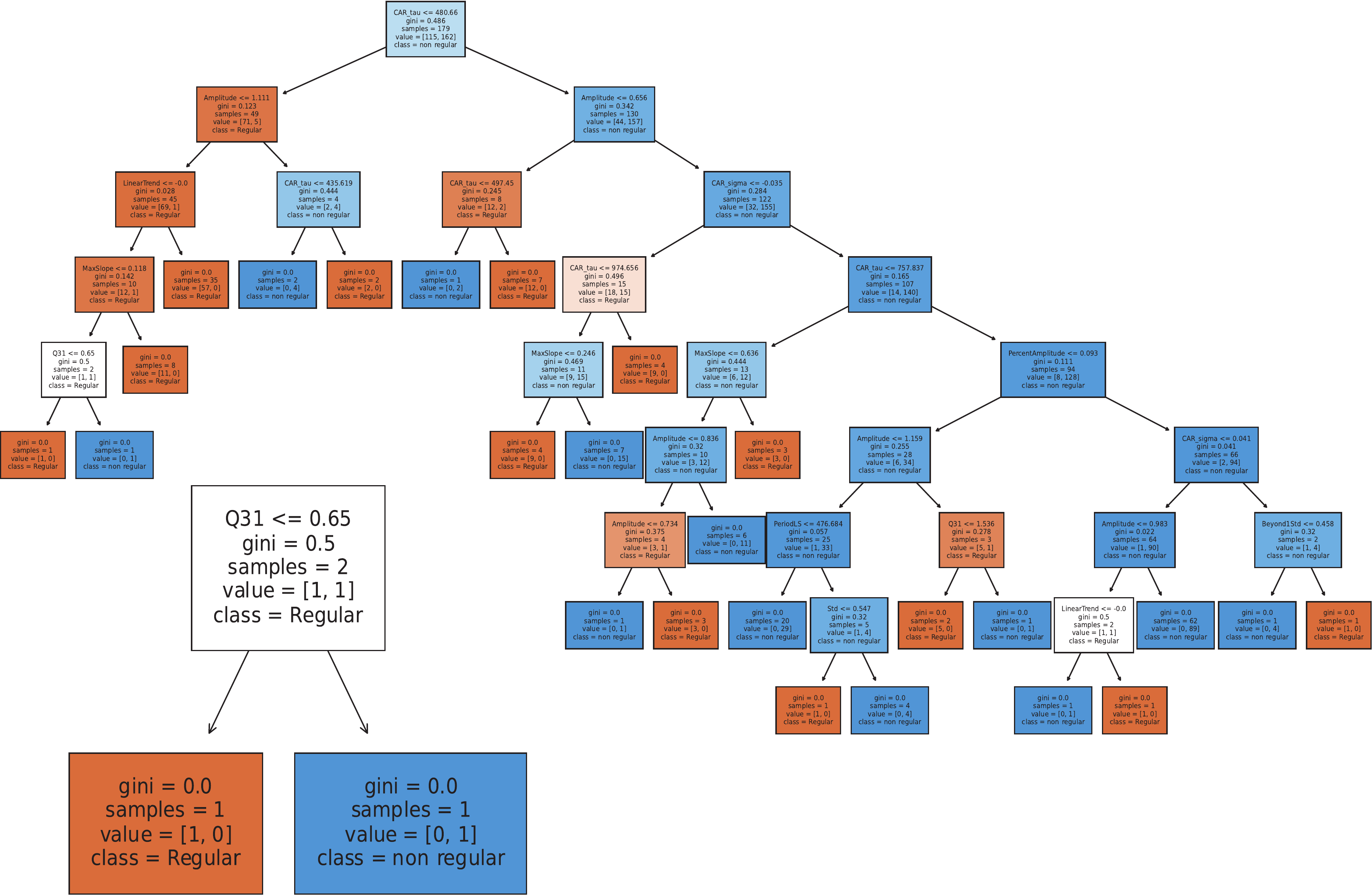}
    \caption{An example of the decision trees for one randomly selected Mira in our sample. An example of the leaves in the decision trees is enlarged and shown on the bottom-left based on the output from the {\tt Scikit-learn} package. Regular and non-regular Miras are indicated in the orange and blue boxes, respectively. The decision tree works as follows: first, the algorithm collects the features in an initial root node (the top cyan box) from all training data set. Then, the RF algorithm divides the decision tree into two leaves based on the features of the input training data, and each leaf is further divided into two leaves until the groups in the leaves are all identical. An example is illustrated in the bottom-left with enlarged boxes (or leaves). The white box represents a leaf with two samples. If one of them have $Q31$ smaller than $0.65$ (based on the training set), then it is classified as a regular Mira in the orange box, else it is classified as a non-regular Mira in the blue box.}
    \label{DT}
\end{figure}


\begin{figure}
    \centering
    \includegraphics[width=1\columnwidth]{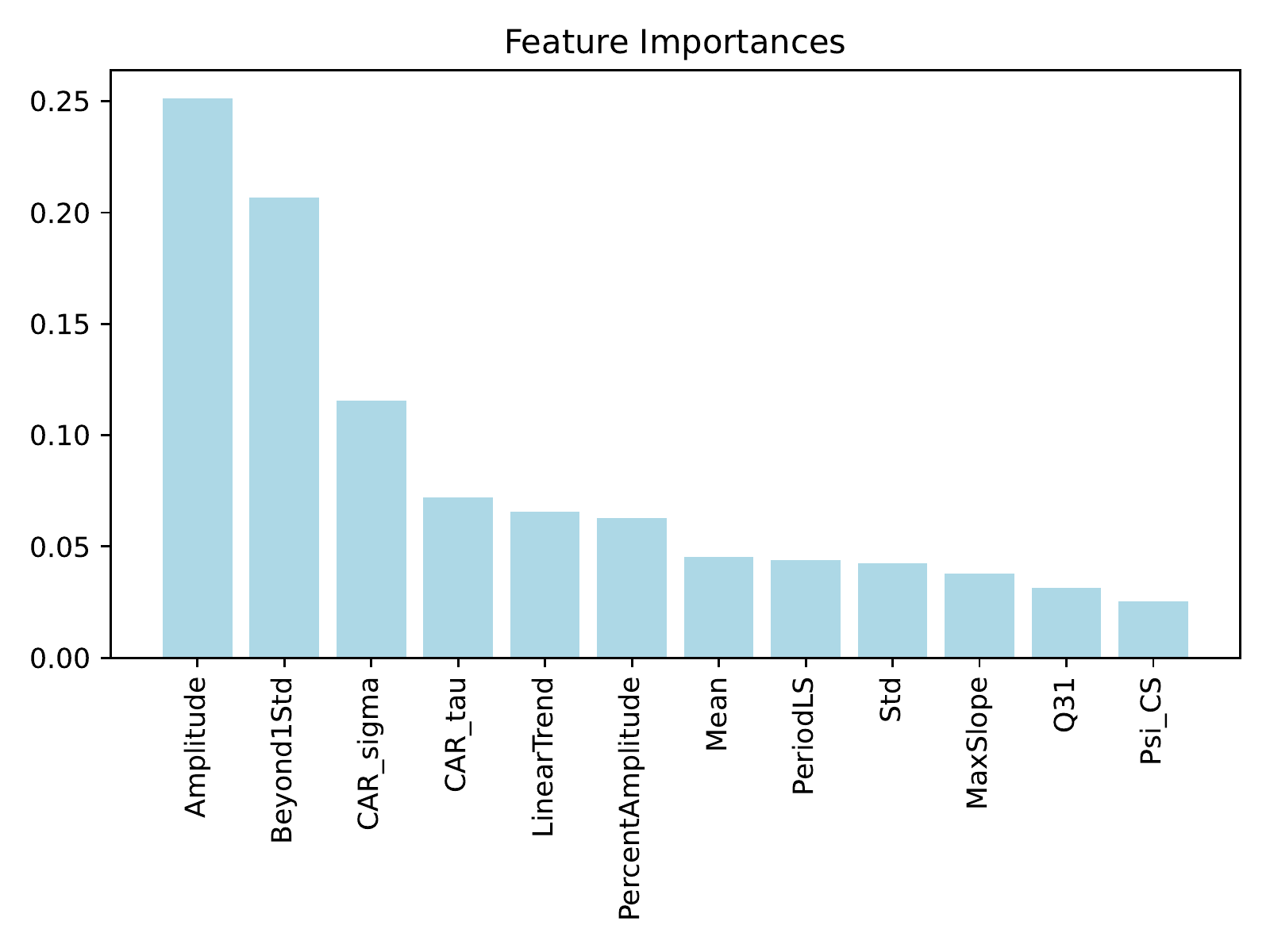}
    \caption{Rank of the importance of the light curve features.}
    \label{FI}
\end{figure}

\begin{figure}
    \centering
    \includegraphics[width=1\columnwidth]{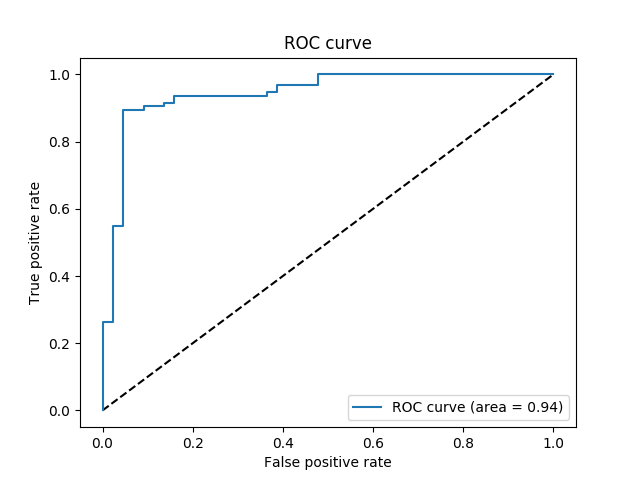}
    \caption{The receiver-operating-characteristic (ROC) curve applied the RF algorithm to our training set. The area under the ROC curve (AUC) represents the performance of RF algorithm; the value of 1 means perfect performance. The dash line indicates AUC = 0.5, implying random guessing. In our case, we achieved an AUC of 0.94.}
    \label{RFRT_LMC}
\end{figure}

\begin{figure}
    \centering
    \includegraphics[width=1\columnwidth]{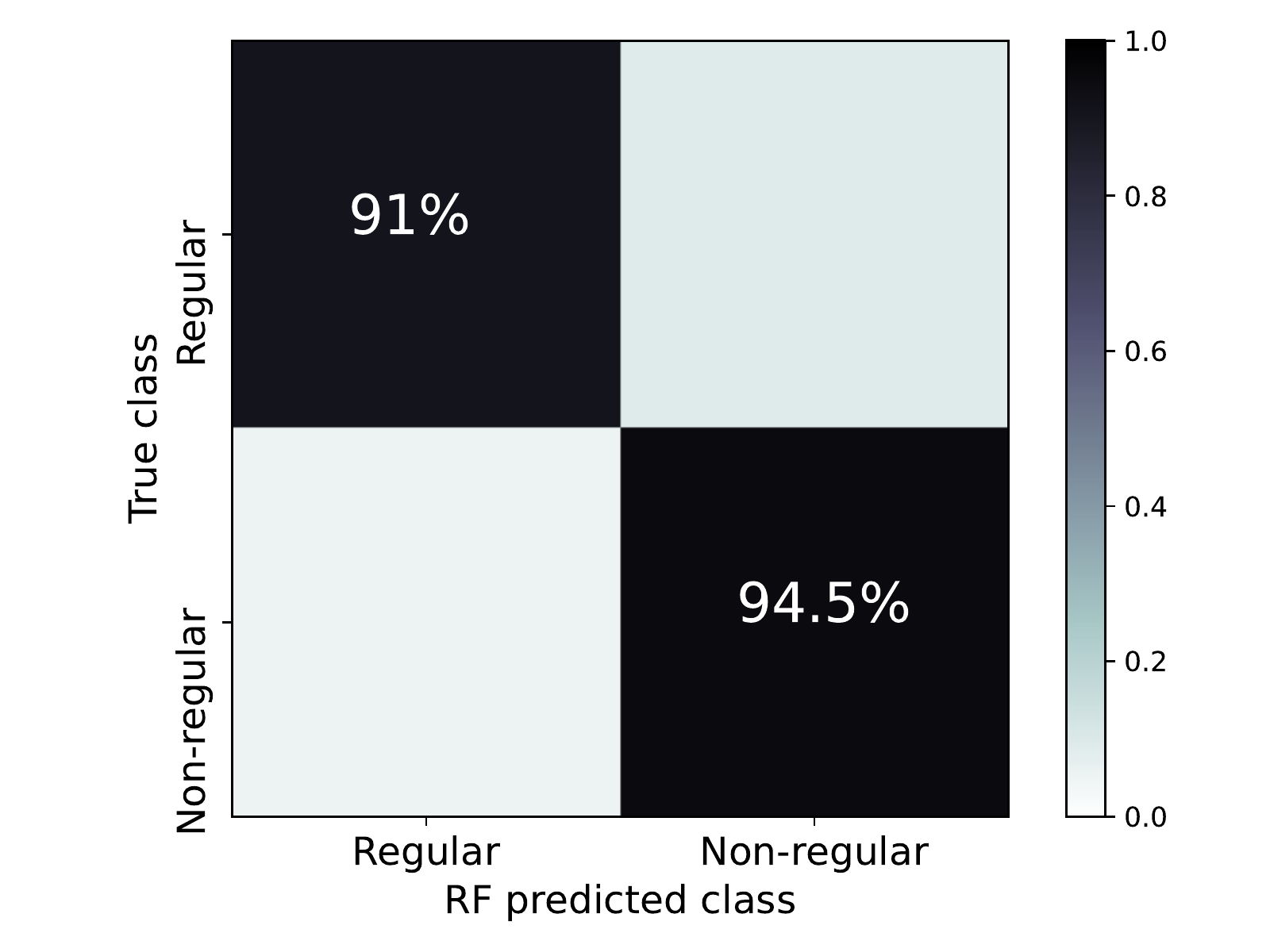}
    \caption{The confusion matrix for the RF classification is based on the test data selected as part of the training set data.}
    \label{CM}
\end{figure}

We retrieved the OGLE-III I-band light curves for 1663 and 352 Miras in LMC and SMC, respectively, classified by the OGLE team \citep{2009AcA,2011AcA}. These photometric data cover the time span between 3000 to 4500 days. Visual inspections of a small subset of these light curves revealed that some Miras are regular Miras with single periods, while other Miras exhibit long-term variation or multi-periodic behaviors as mentioned in the Introduction (see Figure \ref{LC_example} for representative examples). To classify the Mira light curves into regular and non-regular Miras, we employed the powerful machine learning (ML) techniques that are becoming popular and widely used in astronomy, especially for classification purposes \citep{Ra15}. We selected the random forest (RF) algorithm as our ML classifier due to its high efficiency, simple use, and high modifiability, which has been applied to a variety of datasets to classify astronomical sources \citep[for example, see][]{BL2001,2021MNRAS.501.3457P}.

The RF algorithm required users to select "features" to perform the classification. In this work, we selected 12 light curve features generated from the Python package {\tt FATS} (Feature Analysis for Time Series, \citealt{nun15,2017ascl.soft11017N}). These light curve features are Amplitude, Beyond1Std, CAR\_sigma, CAR\_tau, Mean, LinearTrend, Percent Amplitude, Period LS, Std, MaxSlope, Q31, and Psi\_CS. Definitions of these features can be found in the {\tt FATS} document\footnote{ \url{ http://isadoranun.github.io/tsfeat/FeaturesDocumentation.html}} and will not be repeated here. After initial visual inspections on a subset of the light curves, we selected 100 and 150 regular and non-regular Miras that appeared significantly distinct to be our training data to ensure we have at least 100 representative light curves for both types of Miras as our training sets. Then we used the RF classification subroutine available from the {\tt Scikit-learn} \citep{JLMR} package to perform the light curves classification.

The RF algorithm was comprised of a large number of decision trees. Each decision tree can classify a selected target with the given input features. An example of such a decision tree was shown in Figure \ref{DT}. The RF algorithm will perform a majority voting based on all decision trees to avoid errors caused by using just a single decision tree. Our RF has 5000 decision trees in our study, and the minimum number of decision leaves was set to 50. Based on our training data, the ranks of the feature importance were displayed in Figure \ref{FI}, at which the three most important features are Amplitude, Beyond1std, and CAR sigma. If some of the features were removed during the training process, we found that accuracy on the classification would decrease. Therefore, we kept all of the 12 features in our RF classification, achieving an accuracy of 93.2\%. The resulting receiver-operating-characteristic (ROC) curve is presented in Figure \ref{RFRT_LMC}, showing that the RF algorithm can perform reasonably well for our classification purpose. We have also used $70\%$ training data to be the test data to check the performance of the RF classification. The result is presented in Figure \ref{CM}, showing that the True-False ratios (TFR) for regular Mira and the non-regular Miras are $91\%$ and $94.5\%$, respectively. Since both values are greater than $90\%$, we believe the RF algorithm classification can be applied to classify the Miras into regular and non-regular Miras based on their light curve features.  

Based on our RF classification, there were 694 (642 in LMC and 52 in SMC) regular Miras and 1321 (1021 in LMC and 300 in SMC) non-regular Miras in our samples. A digital table that included the numerical values for all of the features for our data (including the training set) and classification results were available upon request.


\section{Regular Mira}

\begin{figure*}
    \centering

    \plottwo{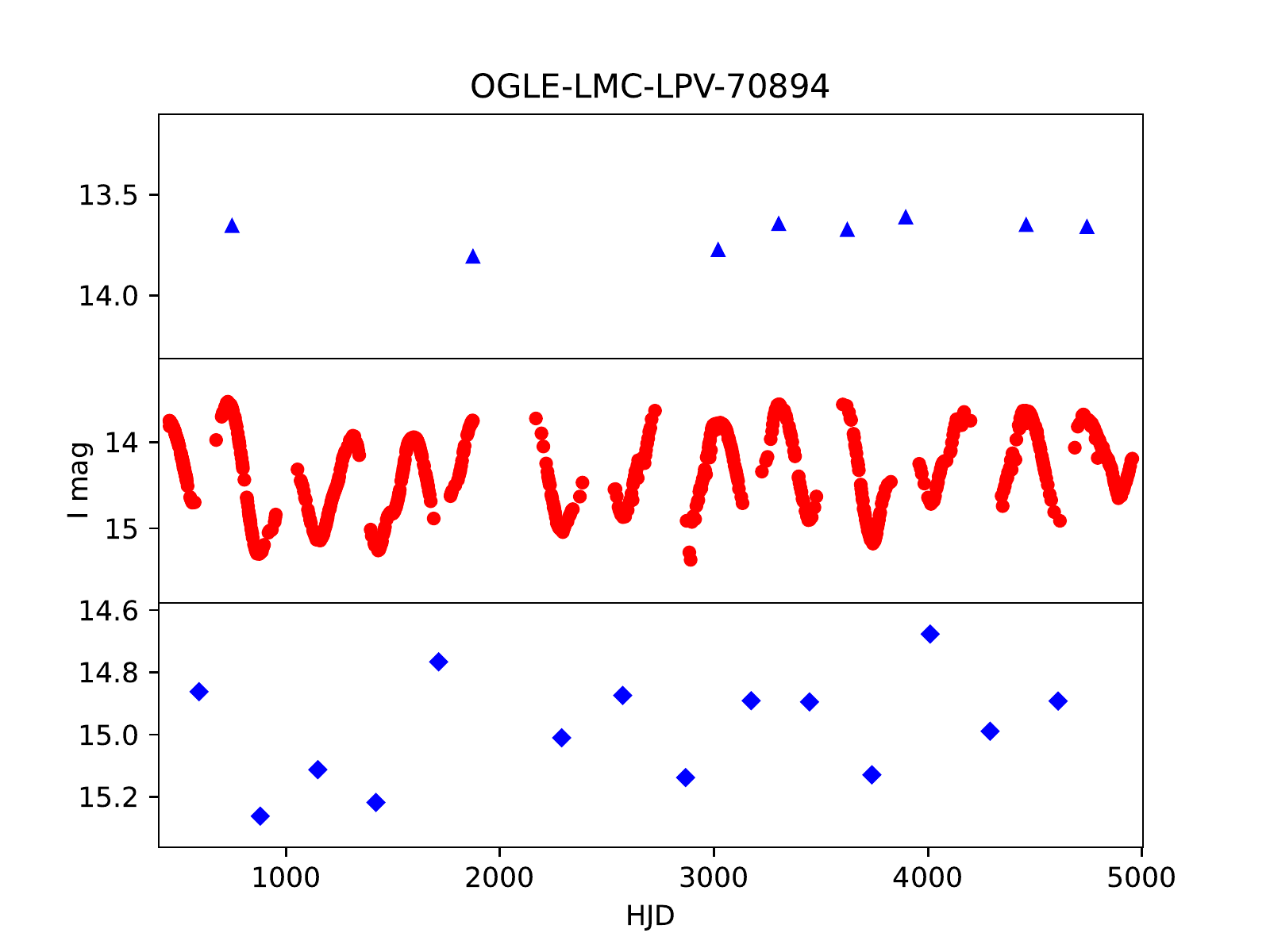}{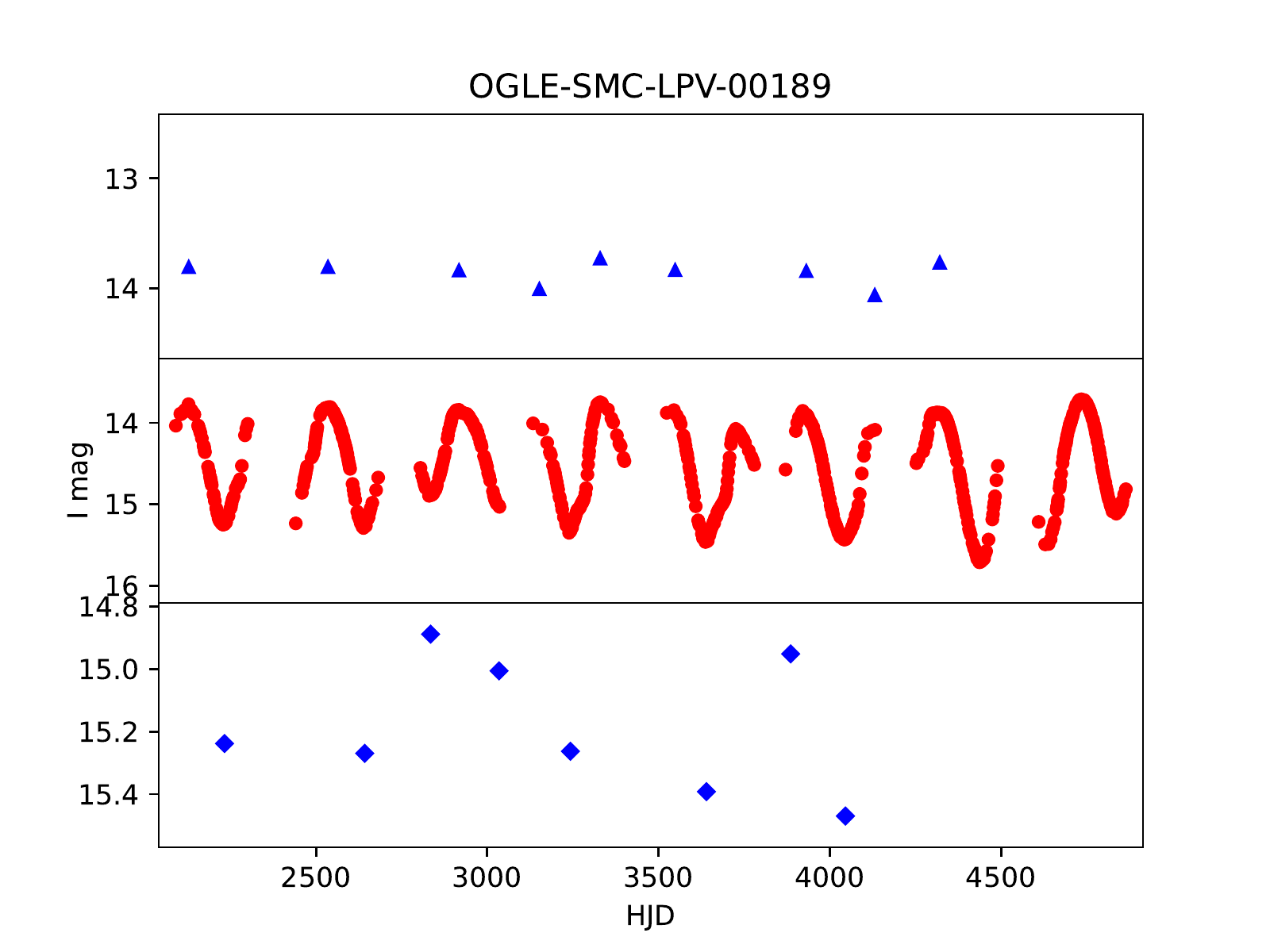}
    \plottwo{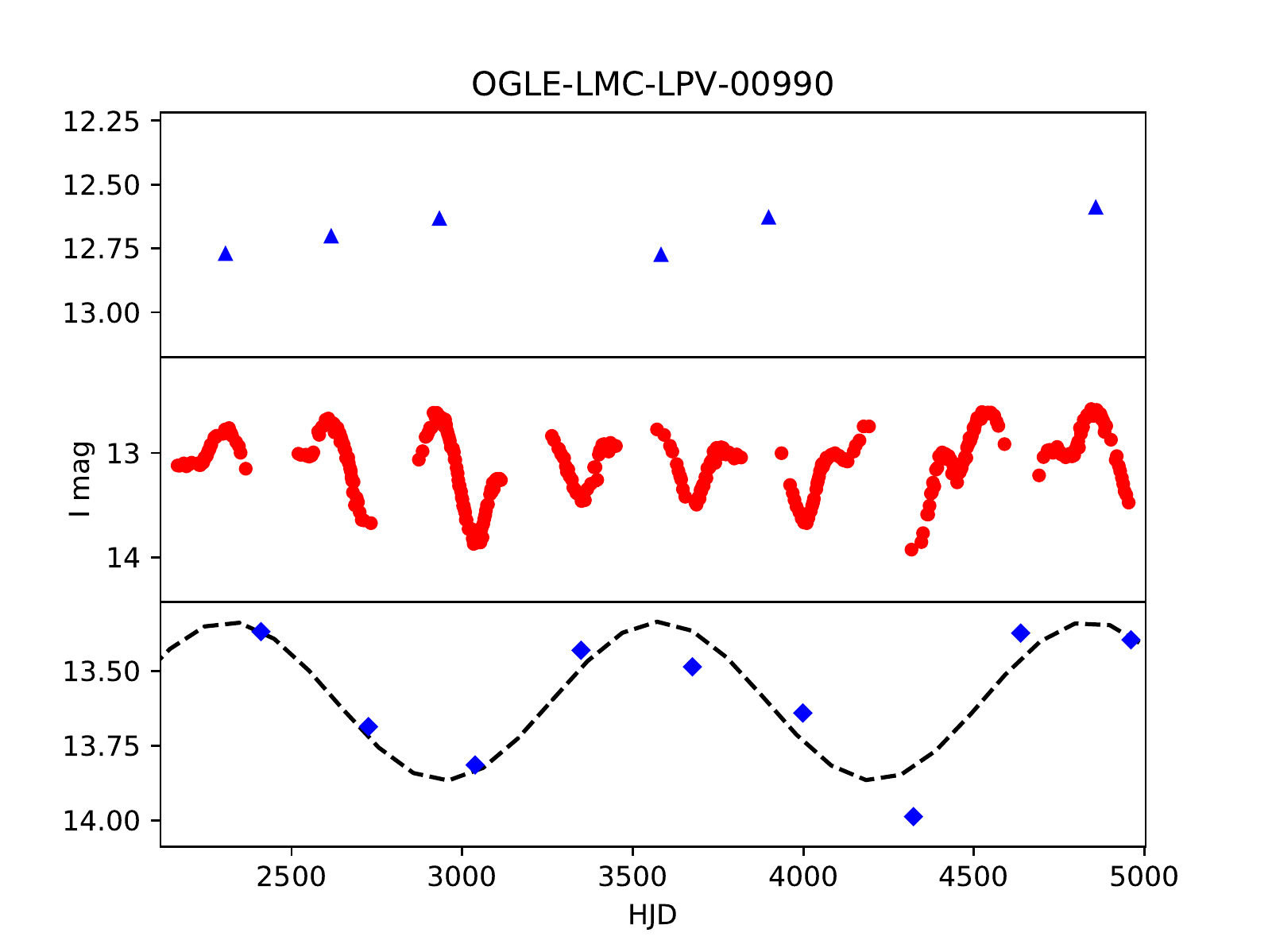}{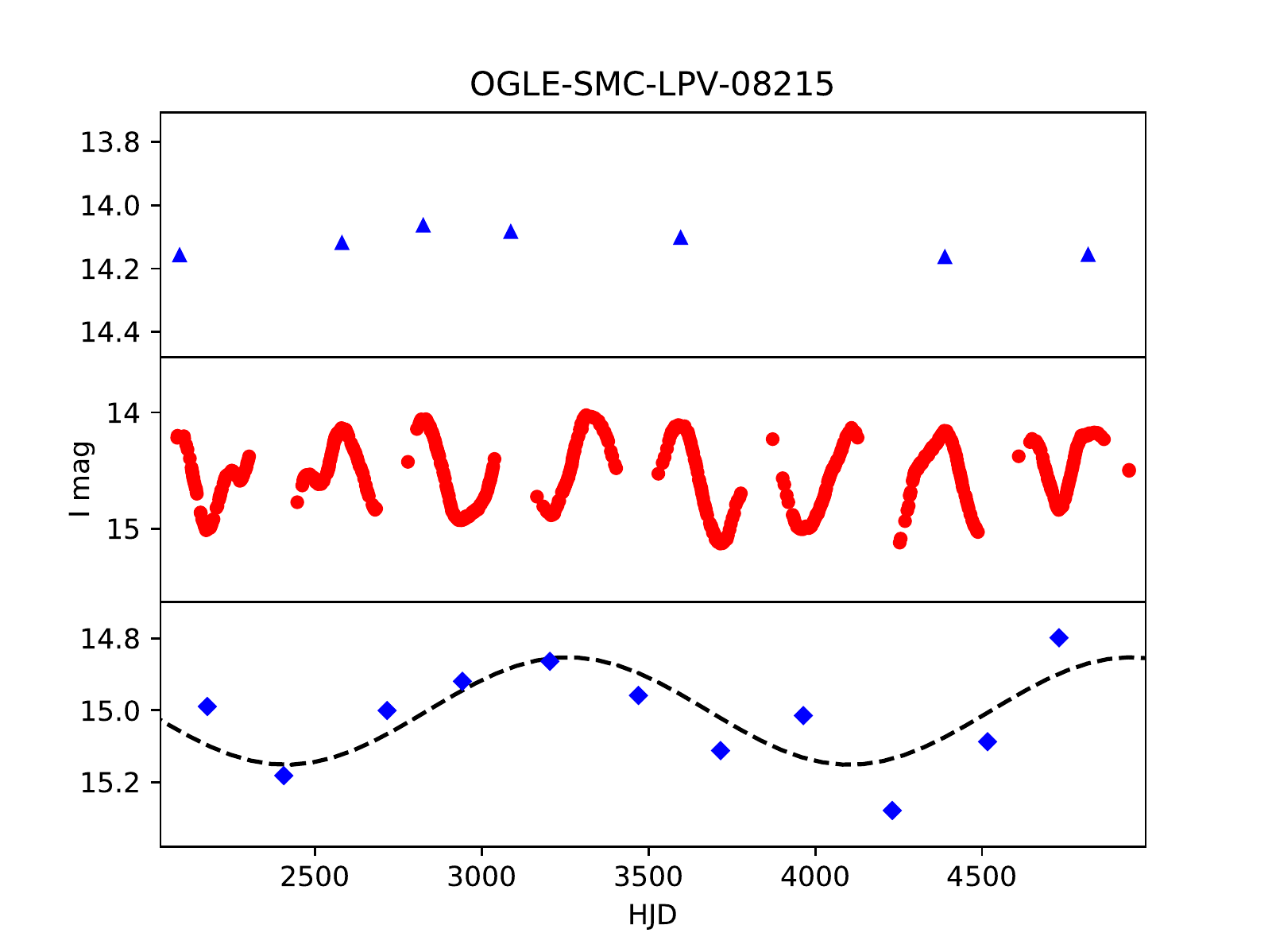}
    \plottwo{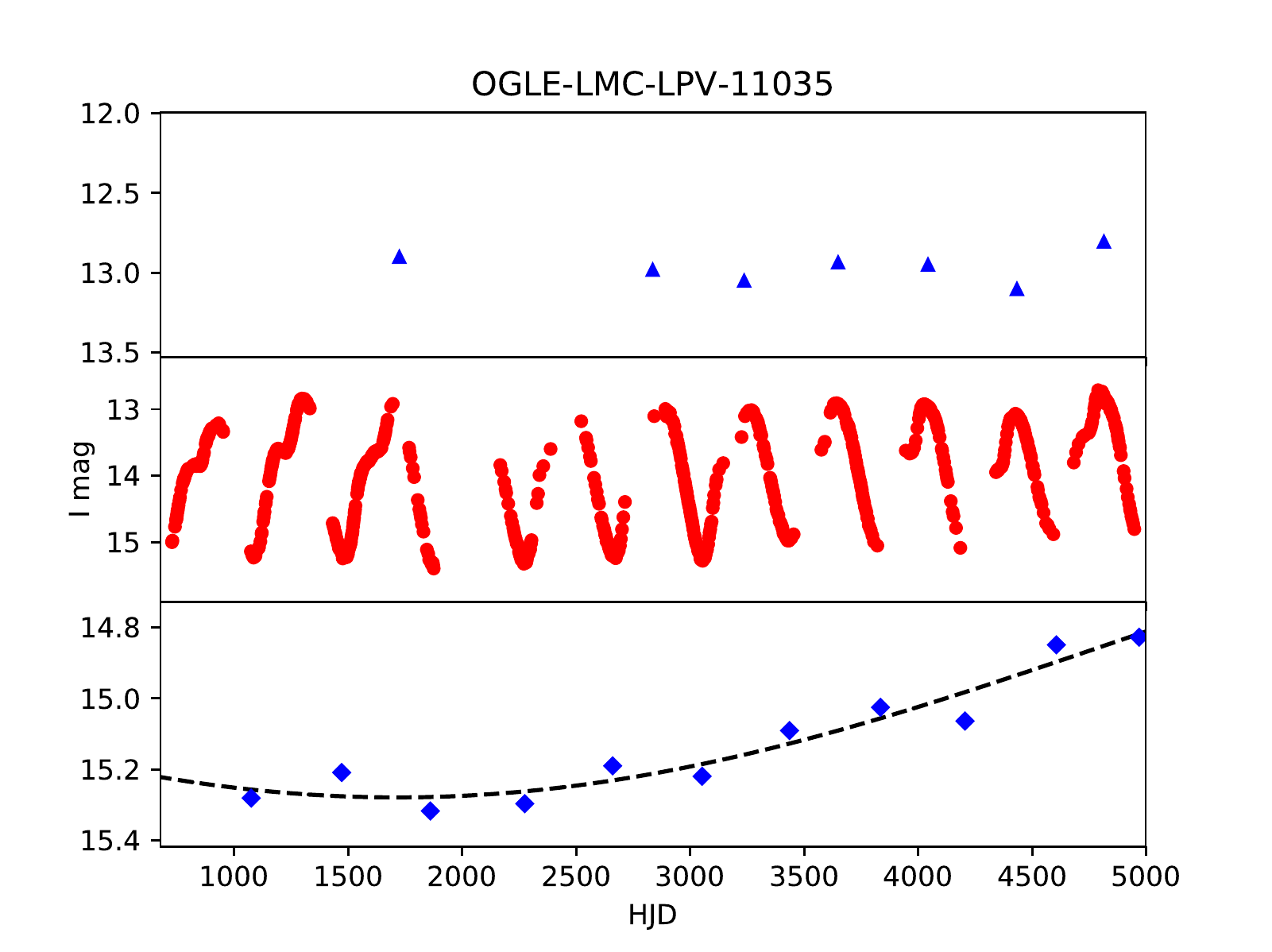}{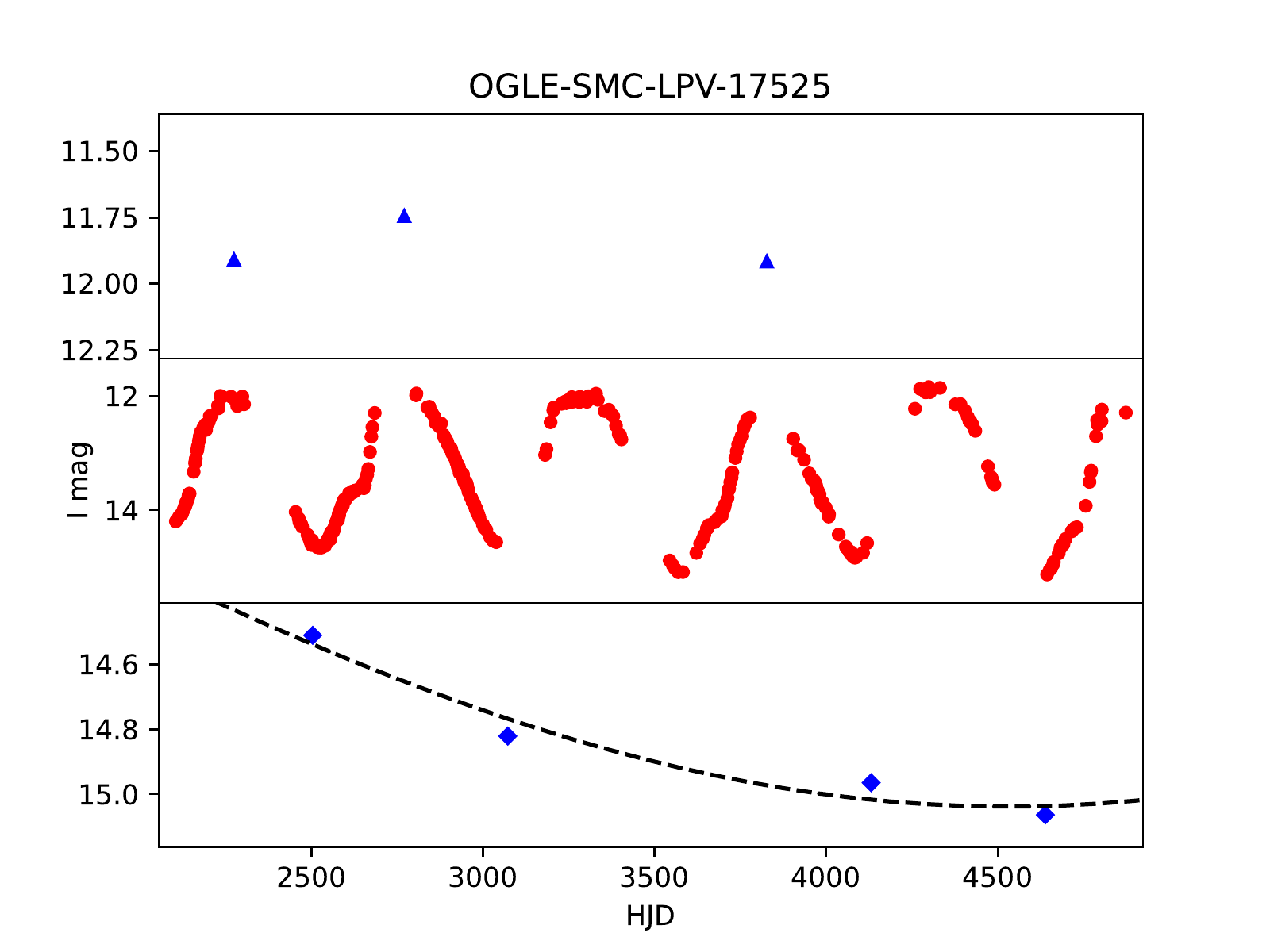}
    \caption{Examples of the light curves for regular Miras in the LMC (sub-figures on the left) and SMC (sub-figures on the right). The middle panel presents the I-band light curve taken from the OGLE-III in each sub-figures. The top and bottom panels show the determined magnitudes at maximum and minimum light, respectively, for each pulsation cycle (whenever available). The top two sub-figures are examples of light curves with minimal light magnitudes that do not exhibit any periodicity. The middle two sub-figures are the light curves such that the magnitude at minimum light can be fitted with a periodic sinusoidal function (shown as dashed curves) with periods $P_{min}$, around 1000 days. The bottom two sub-figures are similar to the sub-figures in the middle, except $P_{min}$ were found to be much longer than 3000~days.}
    \label{max}
\end{figure*}

\begin{figure*}
    \centering
    \includegraphics[width=1\columnwidth]{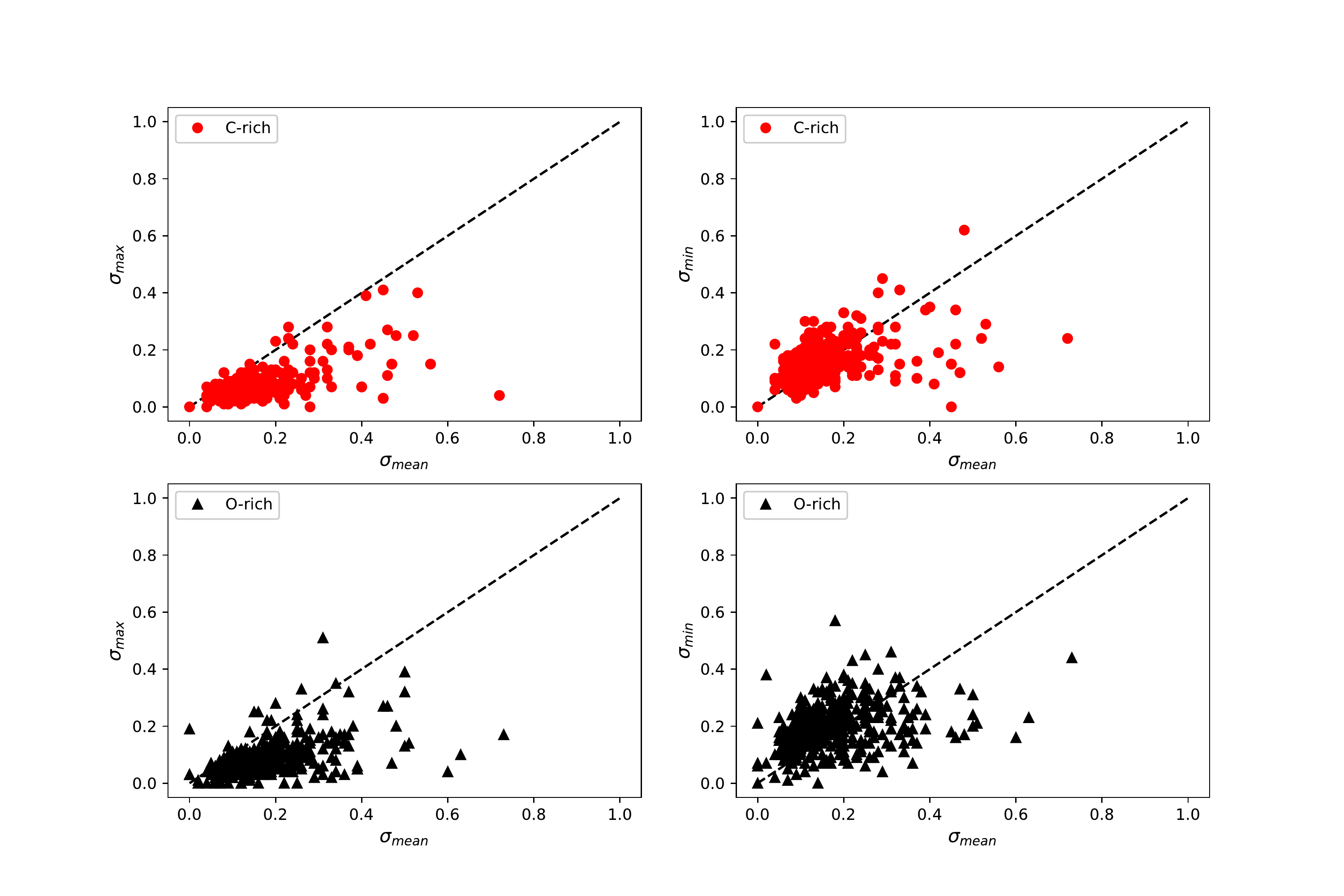}
    \caption{Correlations of the standard deviations $\sigma$ for the maximum, mean, and minimum magnitudes. The left panels are for $\sigma_{mean}$ vs. $\sigma_{max}$, and the right panels show the $\sigma_{mean}$ vs. $\sigma_{min}$, separated for the C-rich (top panels) and O-rich (bottom panels) regular Miras. The dashed lines indicate a 1:1 correlation.}
    \label{sigmap}
\end{figure*}

Based on the results found in Section 2, the RF algorithm classified 642 (232 C-rich Miras and 410 O-rich Mira) and 52 (38 C-rich and 14 O-rich) regular Miras in the LMC and SMC, respectively. Light curves of these regular Miras shared similar patterns with sinusoidal variations; few examples are illustrated in the middle panels of the sub-figures in Figure \ref{max}. Therefore we used the Lomb-Scargle (LS) periodogram, implemented in {\tt astropy} \citep{astropy2013,astropy2018}, to calculate their periods. 
One common pattern for the light curves of regular Miras was that their magnitudes corresponding to the maximum light were very stable. While at minimum light,they  displayed a large fluctuation. 

 We first divided the light curve for the regular Miras into individual pulsation cycles based on the computed LS period. For each pulsation cycle, if the number of data points was more than 15, the data set were fitted with a sinusoidal curve and then the curve was used to obtain the maximum, mean, and minimum magnitudes. Finally, the averages and standard deviations of the maximum, mean, and minimum magnitudes were calculated by combining the cycles' values. Upper and lower panels in each sub-figures of Figure \ref{max} present the stable magnitudes at maximum light and unstable magnitudes at minimum light, respectively. Figure \ref{sigmap} presents the correlation between the various standard deviation ($\sigma$). According to Figure \ref{sigmap}, the $\sigma_{mean}$ were correlated with the $\sigma_{min}$, in contrast the $\sigma_{max}$ are in general smaller than the $\sigma_{mean}$. Therefore, we can see that the values for $\langle m_\mathrm{max}\rangle$ are more stable than $\langle m_\mathrm{mean}\rangle$.

Interestingly, the magnitudes at minimum light for some regular Miras were found to exhibit variations with time. Some variations seem to be periodic, so we tried to find the possible periods. We performed a period search using trial periods in three ranges: those smaller than 3000 days, between 3000 and 10000 days, and larger than 10000 days. After the best periods ( $P_{min}$) were found, a simple sinusoidal function was used to fit the magnitudes at minimum light (dashed curves in Figure \ref{max}) and a standard deviation of residuals value was calculated. If the standard deviation of the residuals value is smaller than 0.04, then the magnitudes at minimum light were considered to be periodic. We split the behavior of magnitudes at minimum light into three different classes: the first class was for the regular Miras with magnitudes at minimum light that did not show regular periodicity; the second class for those exhibiting periodicity in the range of 300 to 3000 days and third class for those much longer than 3000 days. The number of regular Miras in the first, second and third classes were 197, 363, and 134, respectively. We suggested that if $P_{min}$ is longer than 3000 days, the magnitudes at minimum light may not be truly periodic, but with some long-term trends, In contrast, if $P_{min}$ was around or greater than 1500 days, then there will be pairs of maximum and minimum points in the observed time span, so we believed that the period should be real.\footnote{We emphasize that we are using the light curves from OGLE-III. Confirming or disproving the very long periods of $P_{min}$ has to wait for the OGLE-IV light curves, which are not publicly available when this paper was written.} In Table \ref{data}, we present the average magnitudes and the corresponding dispersion for the available pulsation cycles. The determined $P_{min}$, whenever available, were also given in Table \ref{data}.

\begin{deluxetable*}{llCCCCCCCCC}
\label{data}
\tabletypesize{\footnotesize}
\tablecaption{Summary of photometric properties for regular Miras.}
\tablecolumns{11}
\tablewidth{0pt}
\tablehead{
	\colhead{MIRA\_ID} & \colhead{Spectra type} & \colhead{Period} & \colhead{$\langle m_\mathrm{max}\rangle$} & \colhead{$\sigma_{max}$} & \colhead{$\langle m_\mathrm{mean}\rangle$}& \colhead{$\sigma_{mean}$}& \colhead{$\langle m_\mathrm{min}\rangle$} & \colhead{$\sigma_{min}$} & \colhead{$P_{min}$} & \colhead{ $E(V-I)$ } \\
	\colhead{} & \colhead{} & \colhead{(day)} & \colhead{(mag.)} & \colhead{(mag.)} & \colhead{(mag.)} & \colhead{(mag.)} & \colhead{(mag.)} & \colhead{(mag.)} & \colhead{(day)} & \colhead{(mag.)} 
          }
\startdata
OGLE-LMC-LPV-00082&O-rich&451.5&13.85&0.14&14.43&0.36&14.65&0.07&0& 0.189 \\
OGLE-LMC-LPV-00115&C-rich&176.1&14.57&0.06&14.99&0.17&15.56&0.28&0& 0.219\\
OGLE-LMC-LPV-00355&O-rich&154.6&13.79&0.04&14.23&0.10&14.71&0.14&1099.9 & 0.147 \\
OGLE-LMC-LPV-00743&O-rich&216.1&13.60&0.08&14.11&0.16&14.83&0.18&2400.7& 0.222\\
OGLE-LMC-LPV-00881&O-rich&120.2&13.84&0.02&14.25&0.06&14.67&0.16&3515.4& 0.169\\
$\cdots$ & $\cdots$ & $\cdots$ & $\cdots$ & $\cdots$ & $\cdots$ & $\cdots$ & $\cdots$ & $\cdots$ & $\cdots$ & $\cdots$ \\
\enddata
\tablecomments{The entire Table is published in its entirety in the machine-readable format. A portion is shown here for guidance regarding its form and content.}
\end{deluxetable*}

Since the mean magnitude lies in between the stable magnitudes at maximum and the unstable magnitudes at minimum light, this implies the PL relations for the regular Miras will have smaller dispersion at maximum light when compared to their counterparts at mean light. The reduction of the dispersion for PL relation at maximum light has been discussed by \citet{1997MNRAS.289..428K} and  \citet{2019ApJ...884...20B}. In \citet{2019ApJ...884...20B}, the authors selected $\sim440$ O-rich Miras and compared their PL dispersion at maximum light and mean light. Indeed, dispersion of the PL relations at maximum light was found to be smaller than their counterparts at mean light. Following \citet{2019ApJ...884...20B}, we applied a two-slopes model and a quadratic model to fit the PL relations for both of the O-rich and C-rich regular Miras in our sample. The adopted two-slopes model is:
\begin{equation}
    \label{twoslope}
    m = a + b_{i}(\log P- \log P_b),
\end{equation}

\noindent where the slope $b_{i}$ is different for the short-period ($P<300$~days) and long-period ($P>300$~days) Miras. While the quadratic model is:
\begin{equation}
    \label{quadratic}
    m = a + b_{1}(\log P-\log P_b) +b_{2}(\log P-\log P_b)^{2},
\end{equation}
\noindent where the break period is adopted as $P_b = 300$~days \citep{2019ApJ...884...20B}. Extinction corrections were done using the optical reddening maps of LMC and SMC \citep[][the values of $E(V-I)$ were listed in the last column of Table \ref{data}]{2021ApJS..252...23S} by adopting an averaged value of $R_V=3.41$ and $R_V=2.74$ \citep{1989ApJ...345..245C} for LMC and SMC, respectively.

\begin{deluxetable*}{lCCCC|lCCCC}
\label{tab2}
\tabletypesize{\footnotesize}
\tablecaption{Parameters of the fitted PL relations for regular Miras.}
\tablecolumns{10}
\tablewidth{0pt}
\tablehead{
    \multicolumn{5}{c}{Two-slopes model} & \multicolumn{5}{c}{Quadratic model} \\
	\colhead{} & \colhead{$a$} & \colhead{ $b_{1}$} & \colhead{$b_{2}$} &\colhead{$\sigma$}  & \colhead{ } & \colhead{$a$} & \colhead{$b_{1}$} & \colhead{$b_{2}$} &\colhead{$\sigma$} 
          }
\startdata
\sidehead{LMC only.}
\cutinhead{C-rich ($N=225$)}
Max & 14.75\pm0.56 & -0.44\pm0.23 &-4.95\pm0.88 &0.312&Max & 13.70\pm0.02 & -1.60\pm0.19 & -6.76\pm1.04 &0.307  \\
Mean& 16.08\pm0.57 & -0.77\pm0.23 &-2.66\pm0.67 &0.284&Mean& 14.19\pm0.02 & -1.26\pm0.18 & -3.38\pm0.96 &0.285  \\ 
\cutinhead{O-rich ($N=406$)}
Max & 18.21\pm0.33 & -2.00\pm0.14 & -4.01\pm0.29 &0.327 &Max & 13.23\pm0.02 & -2.70\pm0.09 & -0.97\pm0.34 &0.323  \\
Mean& 16.04\pm0.39 & -0.78\pm0.17 & -2.63\pm0.42 &0.401 &Mean& 14.01\pm0.02 & -1.45\pm0.11 & -0.97\pm0.43 &0.403  \\ 
\cutinhead{All ($N=631$)}
Max & 16.17\pm0.26 & -1.08\pm0.11 & -5.10\pm0.51 &0.334 &Max & 13.46\pm0.01 & -2.31\pm0.09 & -1.67\pm0.35&0.371 \\
Mean& 15.68\pm0.27 & -0.61\pm0.11 & -2.56\pm0.52 &0.372 &Mean& 14.10\pm0.01 & -1.33\pm0.09 & -1.67\pm0.36&0.372 \\
\hline
\sidehead{LMC and SMC combined.}
\cutinhead{C-rich ($N=240$)}
Max & -4.61\pm0.89 & -0.26\pm0.37 & -3.25\pm0.38 &0.337&Max & -5.27\pm0.02 & -1.73\pm0.20 & -7.53\pm1.09 &0.326  \\
Mean& -3.60\pm0.85 & -0.45\pm0.35 & -2.31\pm0.30 &0.293&Mean& -4.75\pm0.02 & -1.38\pm0.18 & -3.66\pm0.97 &0.289  \\ 
\cutinhead{O-rich ($N=419$)}
Max & -1.18\pm0.35 & -1.81\pm0.15 & -3.80\pm0.23 &0.325 &Max & -5.78\pm0.02 & -2.75\pm0.09 & -1.01\pm0.35&0.329  \\
Mean& -3.02\pm0.41 & -0.74\pm0.18 & -2.44\pm0.31 &0.401 &Mean& -4.96\pm0.02 & -1.52\pm0.11 & -1.01\pm0.43&0.405  \\ 
\cutinhead{All ($N=659$)}
Max & -2.80\pm0.31 & -1.08\pm0.13 & -3.92\pm0.22 &0.370 &Max & -5.52\pm0.01 & -2.35\pm0.09 & -1.86\pm0.36 &0.387 \\
Mean& -3.60\pm0.33 & -0.48\pm0.14 & -2.53\pm0.21 &0.370 &Mean& -4.86\pm0.01 & -1.40\pm0.09 & -1.86\pm0.36 &0.375 \\
\enddata
\end{deluxetable*}

\begin{figure}
    \centering
    \includegraphics[width=1\columnwidth]{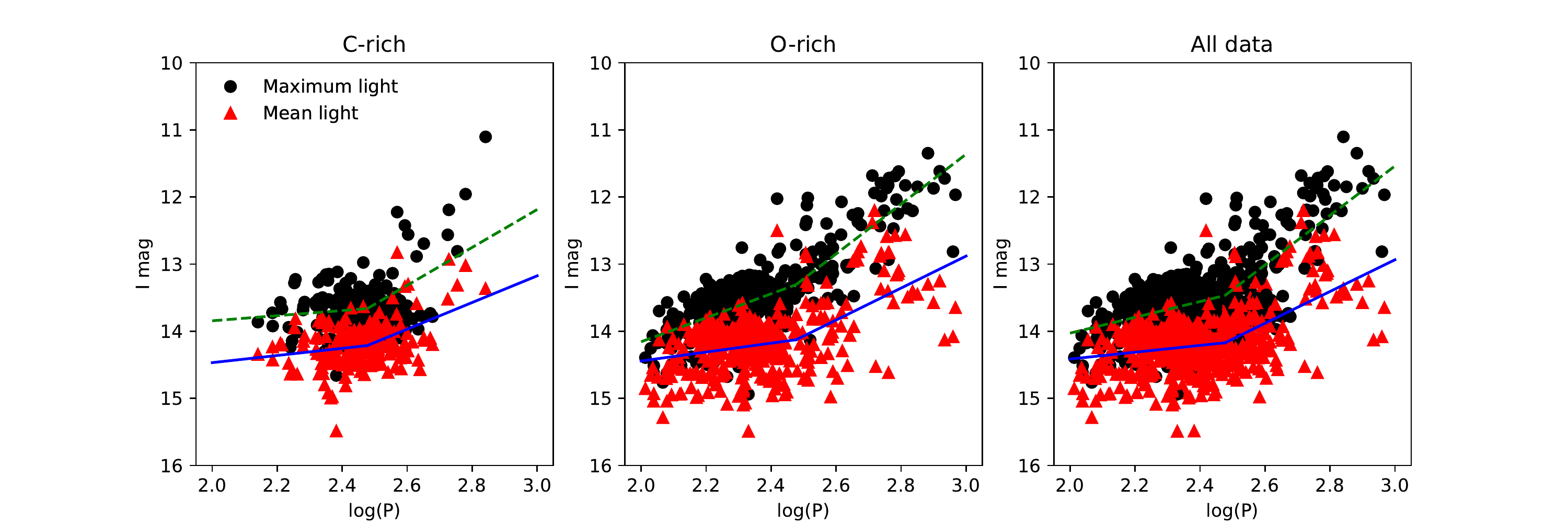}
    \caption{PL relations for regular Miras in the LMC. The green dashed lines and the blue solid lines are the fitted PL relations using a two-slopes model at maximum (black circles) and mean (red triangles) light, respectively.}
    \label{PL_1}
\end{figure}

\begin{figure}
    \centering
    \includegraphics[width=1\columnwidth]{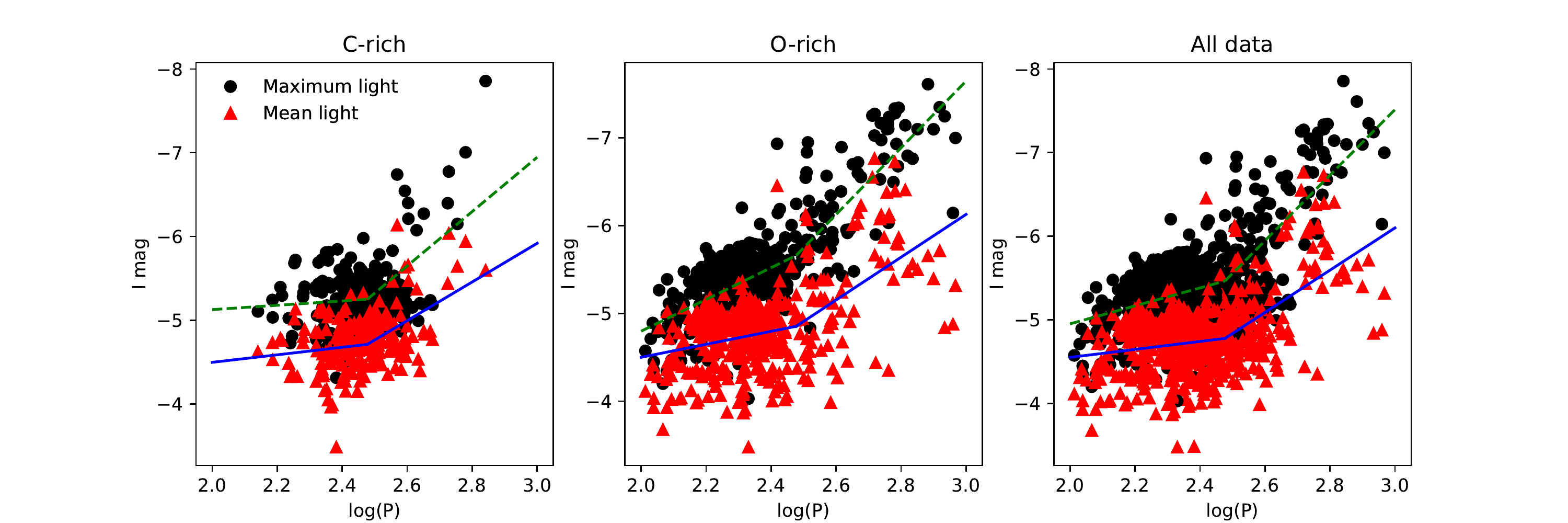}
    \caption{Same as in Figure \ref{PL_1}, but for the combined samples of regular Miras in the LMC and SMC.}
    \label{PL_1_tw}
\end{figure}

\begin{figure}
    \centering
    \includegraphics[width=1\columnwidth]{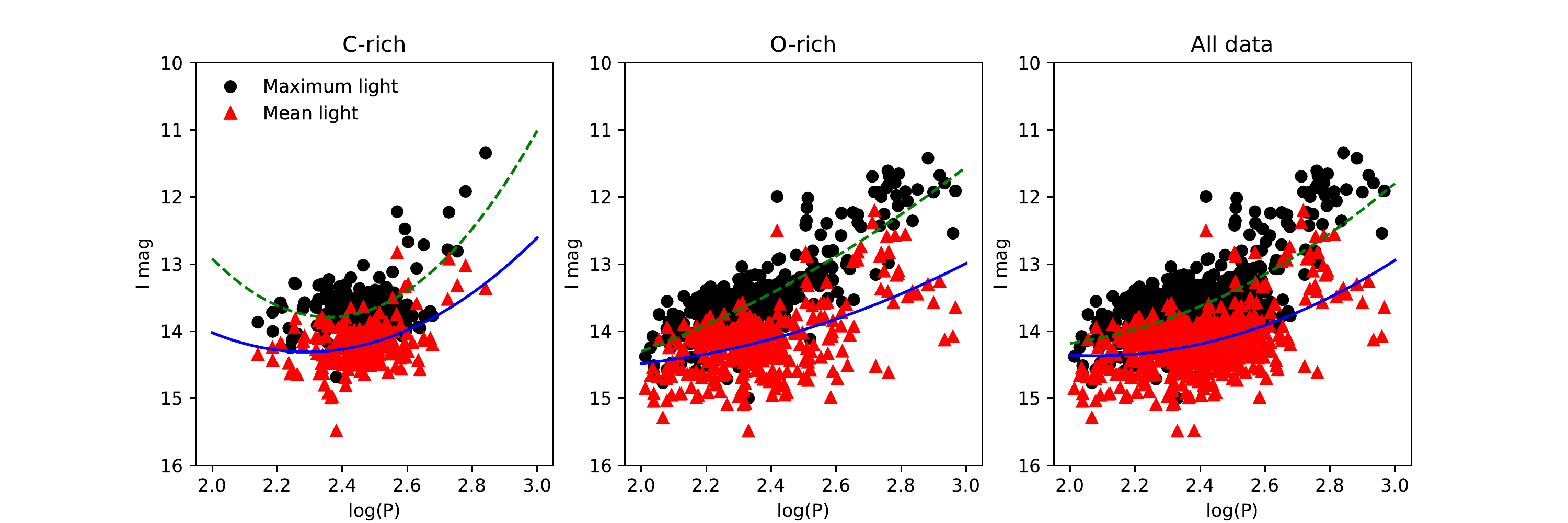}
    \caption{PL relations for regular Miras in the LMC. The green dashed lines and the blue solid lines are the fitted PL relations using a quadratic model at maximum (black circles) and mean (red triangles) light, respectively.}
    \label{PL_2}
\end{figure}

\begin{figure}
    \centering
    \includegraphics[width=1\columnwidth]{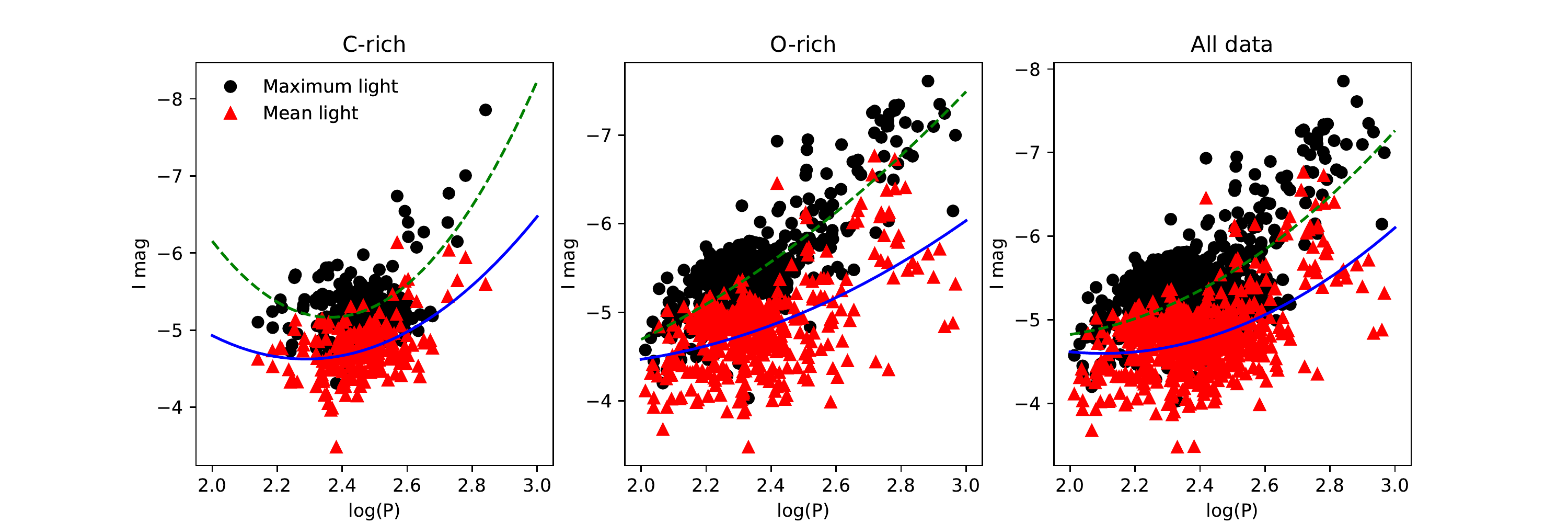}
    \caption{Same as in Figure \ref{PL_2}, but for the combined samples of regular Miras in the LMC and SMC.}
    \label{PL_2_q}
\end{figure}

We fitted these two models of the PL relations using maximum and mean light for regular Miras in two sample sets: one with LMC samples and another with samples combining LMC and SMC. We adopted a distance of $49.59$~kpc \citep{2019Natur.567..200P} and  $62.44$~kpc \citep{2020ApJ...904...13G} for LMC and SMC, respectively, when combining the samples of regular Miras in the Magellanic Clouds. Parameters of the fitted PL relations were summarized in Table \ref{tab2}, while Figures \ref{PL_1} and \ref{PL_2} present the PL relations for the regular Miras in LMC only, fitted with two-slopes and quadratic models, respectively. Similarly, the PL relations for the combined sample for LMC and SMC were also fitted to increase the number of data points in the sample. The results were shown in Figure \ref{PL_1_tw} and \ref{PL_2_q}. 

It can be seen from Table \ref{tab2} that the reduction of the PL dispersion at maximum light only occurs for O-rich regular Miras, either in the sample from LMC only and the sample that combines the LMC and SMC. In \citet{2011MNRAS.412.2345I}, the C-rich Miras did not exhibit a linear relation in the $I$-band with their primary pulsation period because they had different levels of circumstellar extinction. In contrast, we found that C-rich regular Miras display mild linear trends in their PL relation. Furthermore, for the C-rich regular Miras, the dispersion of the PL relation at the maximum light was larger than the dispersion at the mean light using both regression models. Hence, PL relations based on the samples combining both O-rich and C-rich regular Miras were mostly influenced by the C-rich Miras. As a result, PL relation dispersion at maximum and mean light were comparable to each other when the sample of O-rich and C-rich regular Miras were combined (see Table \ref{tab2}). Finally, we pointed out that the PL relation for the C-rich regular Miras at maximum light displays a steep and "up-turn" slope, as demonstrated in Figure \ref{PL_1}, for a $P_b \sim 350$~days. This is because the C-rich Miras with period longer than 350~days have a larger amplitude, hence the magnitudes at maximum light display a larger variation than their shorter period counterparts.



\section{Non-regular Miras}

\begin{figure*}
    \centering
    \plottwo{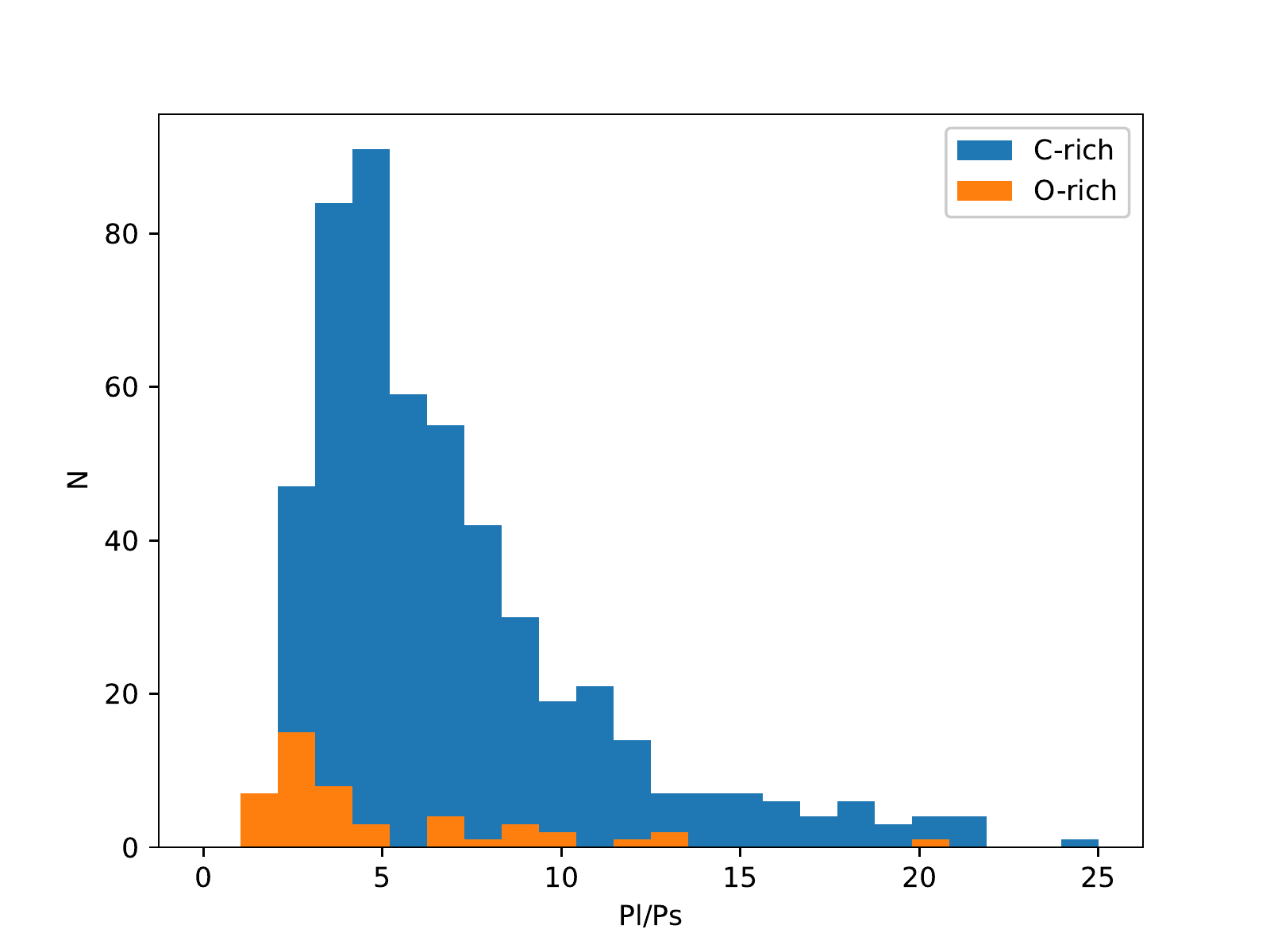}{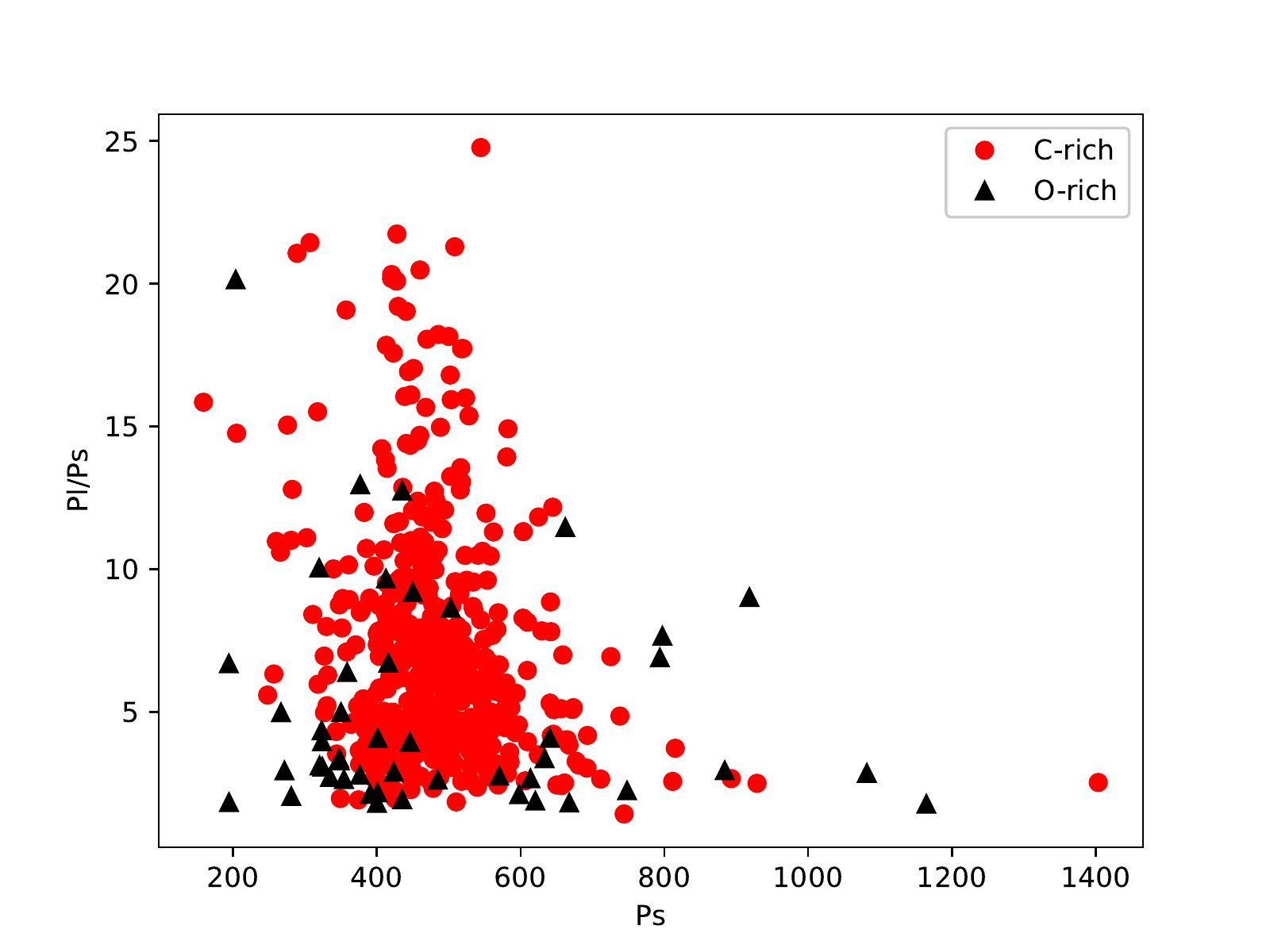}
    \caption{ Left Panel: Histogram of $P_l/P_s$ ratio for non-regular Miras in the sample. Right Panel: The $P_l/P_s$ ratios as a function of $P_s$.}
    \label{hist_multi}
\end{figure*}

\begin{figure*}
    \centering
    \plottwo{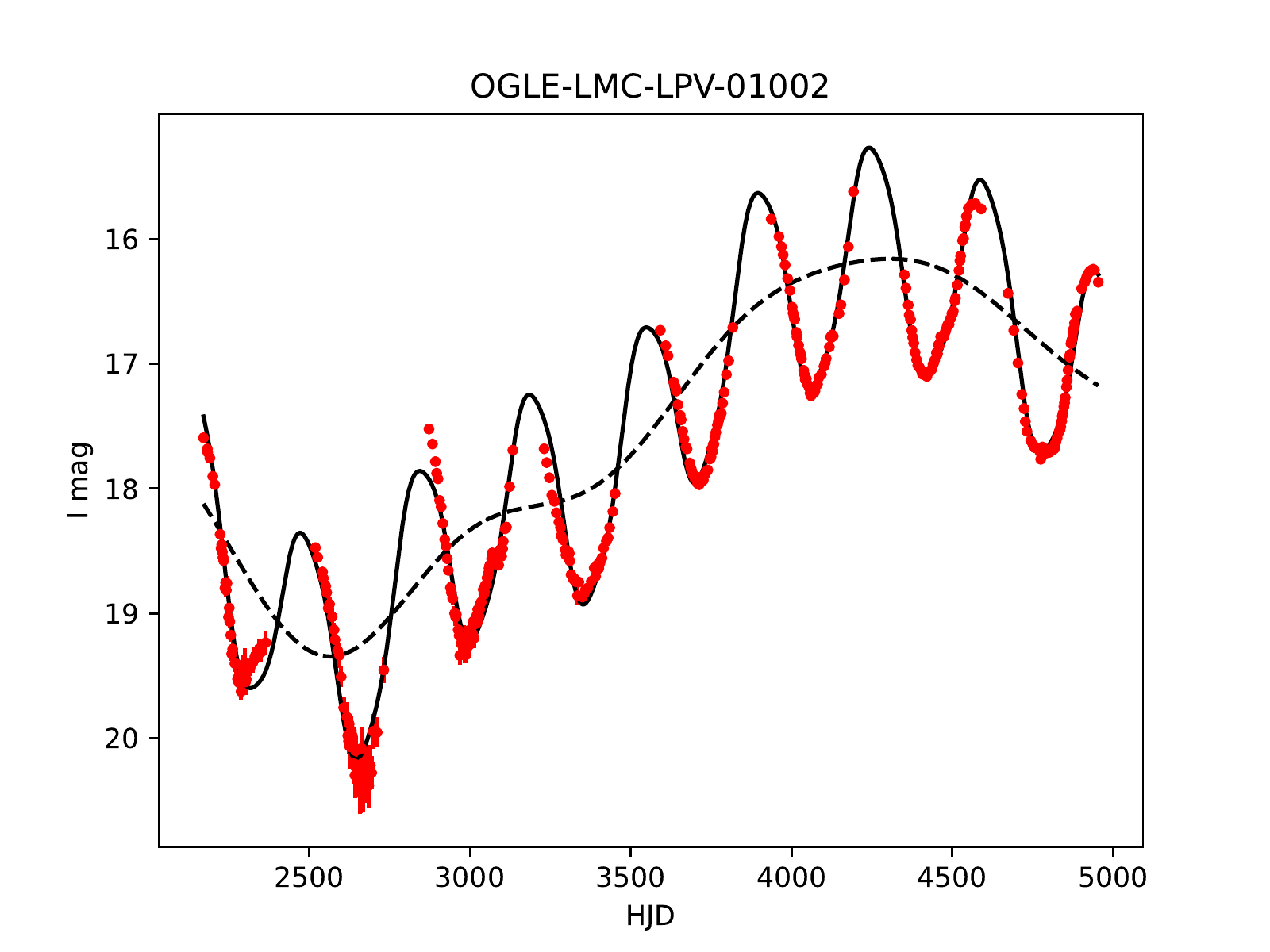}{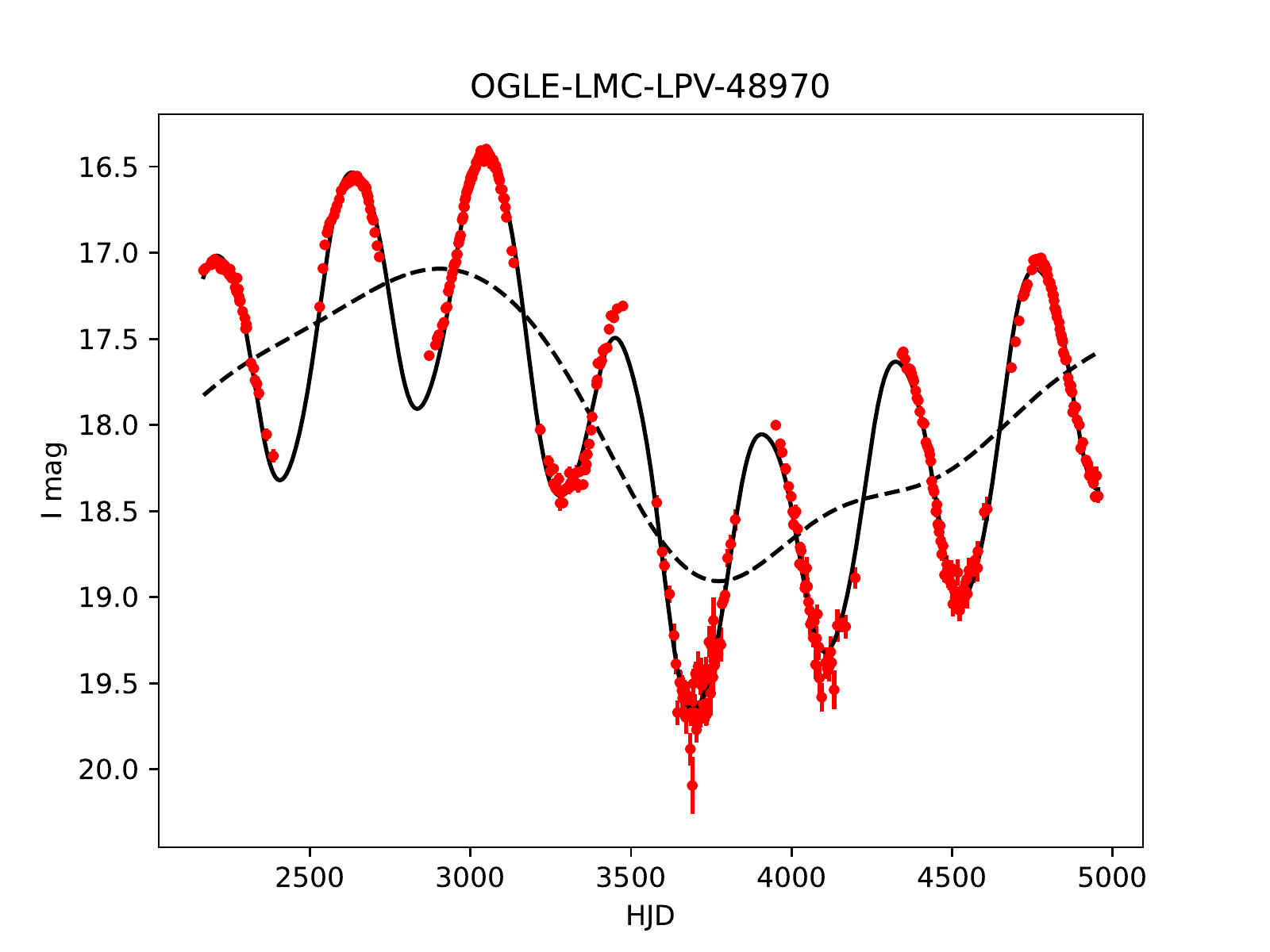}
    \plottwo{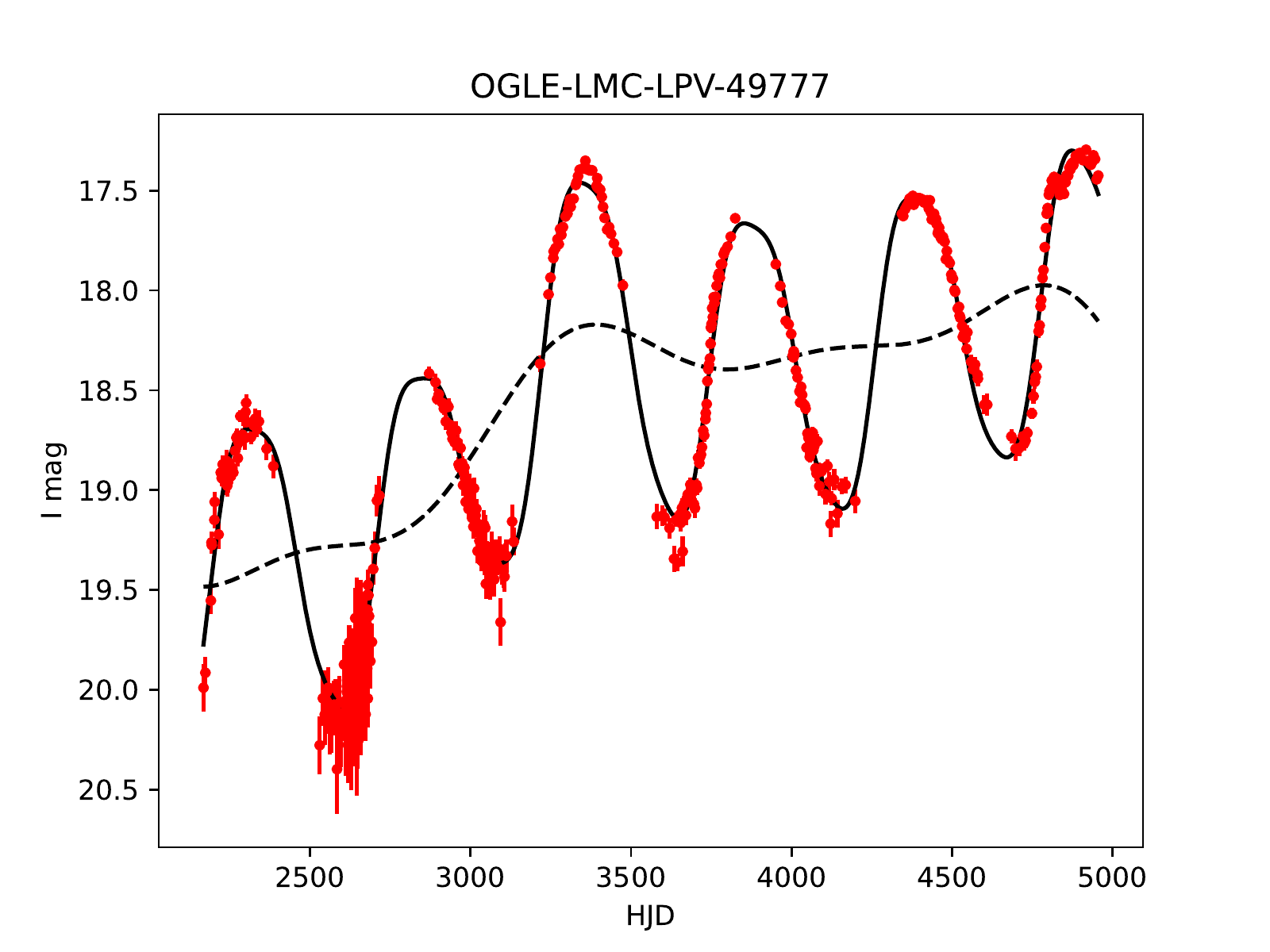}{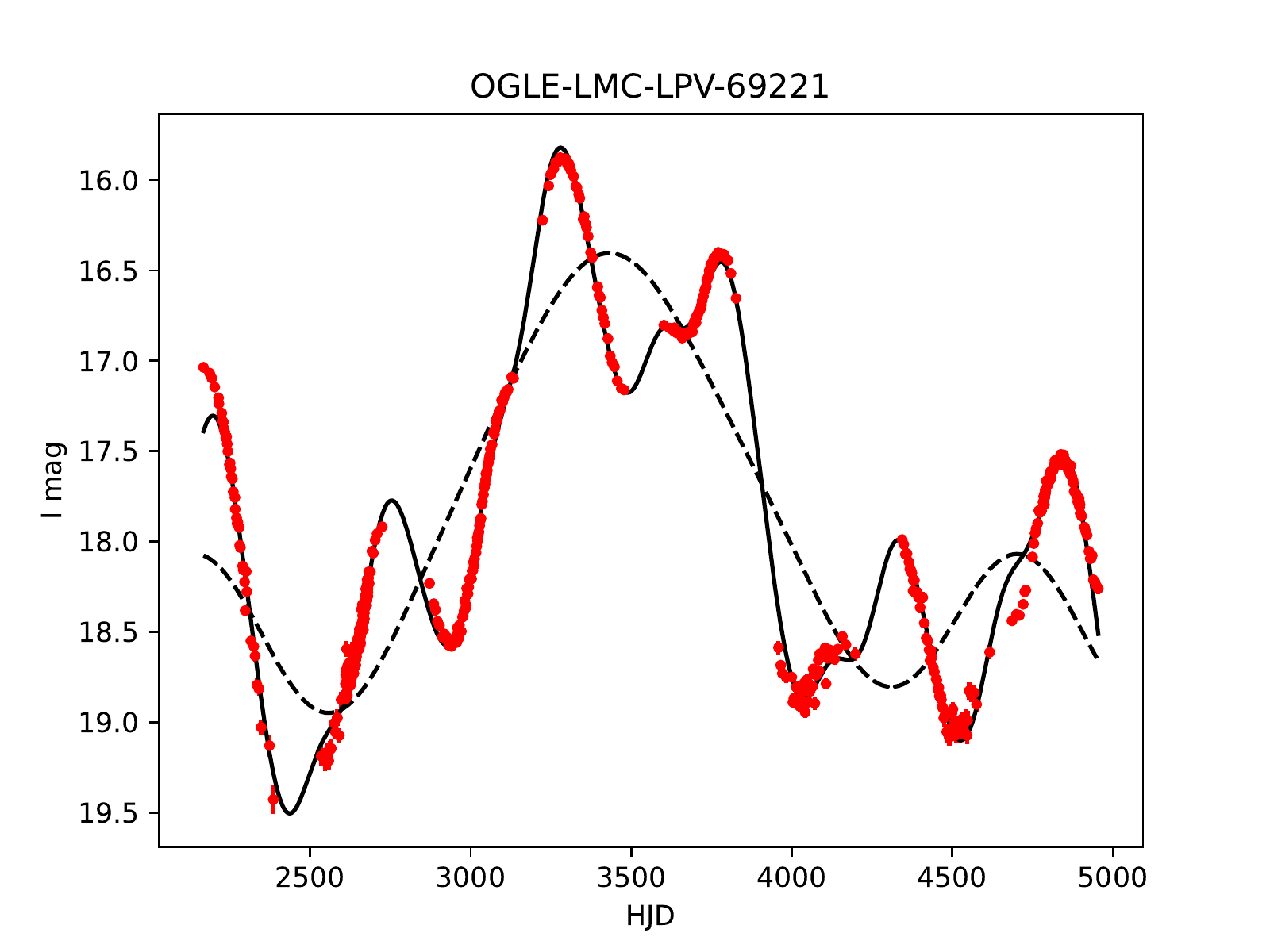}    
    \caption{Examples of light curves for non-regular Miras. The black curves are the best-fit light curves based on the determined short- and long-term periods as described in the text. The dashed curves are the long-term variation $L^l(t)$ given in Equation (3).}
    \label{lc_multi}
\end{figure*}

\begin{figure*}
    \centering
    \begin{tabular}{ccc}
    \includegraphics[width=0.31\columnwidth]{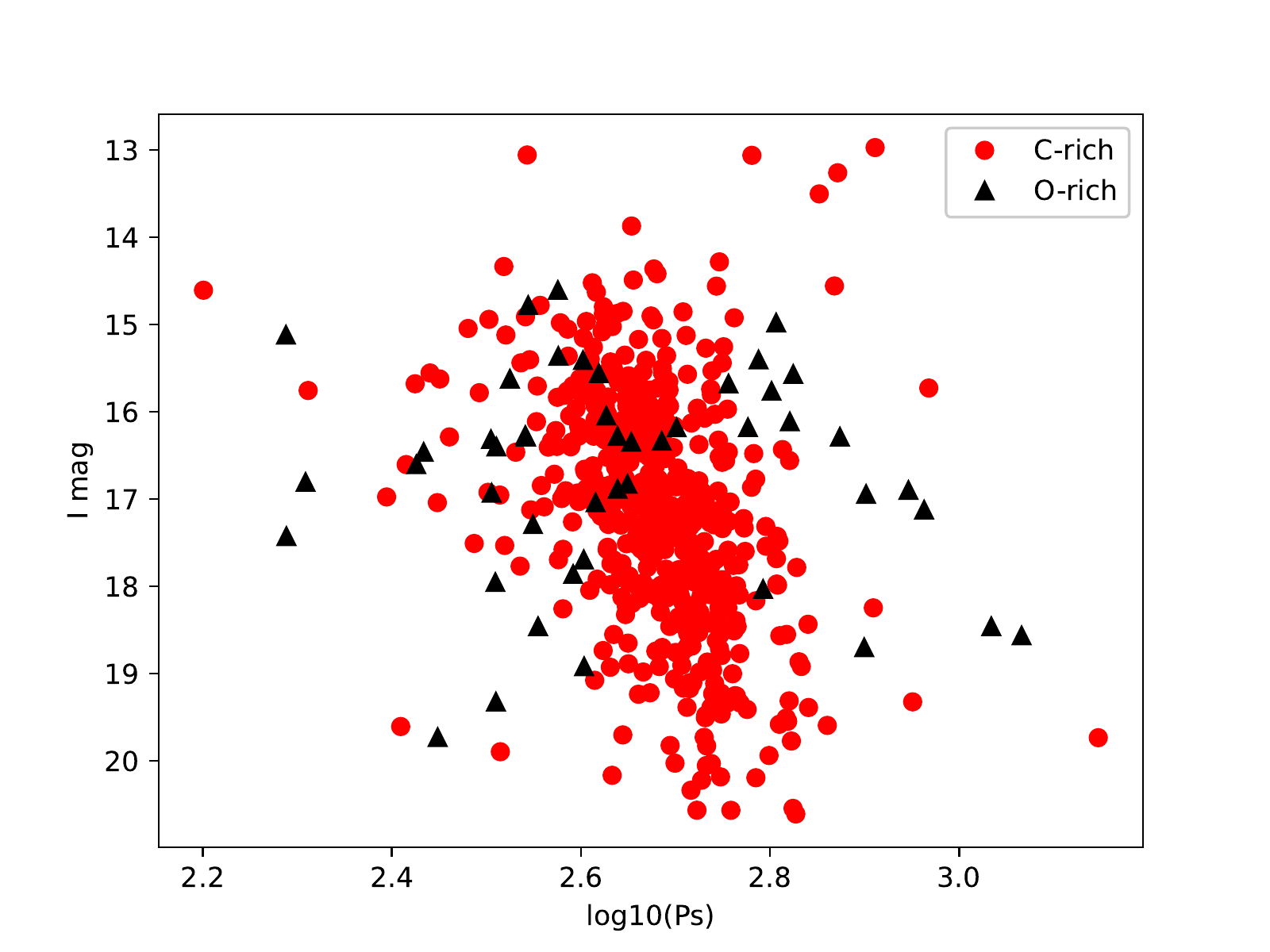} & \includegraphics[width=0.31\columnwidth]{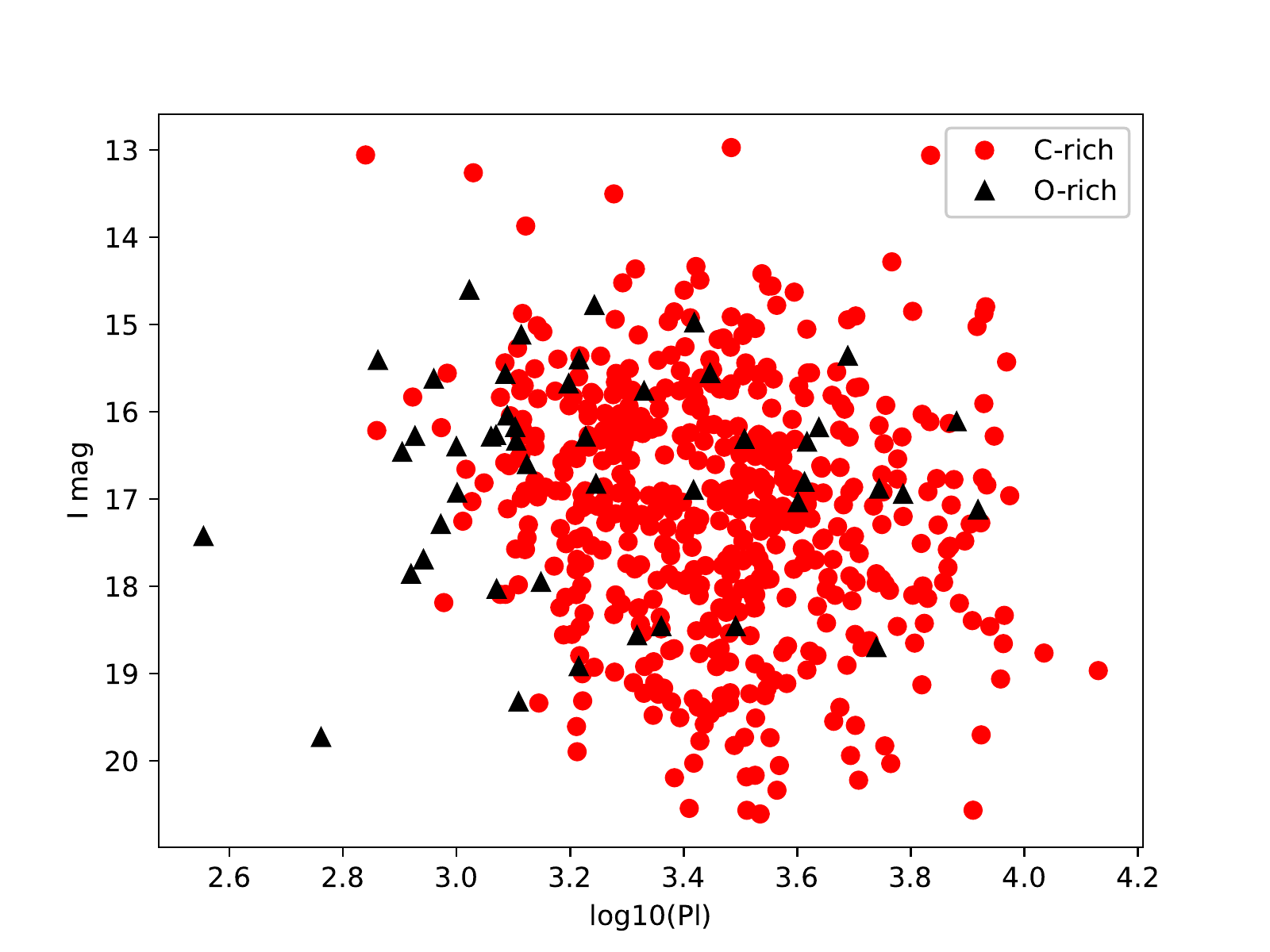} & \includegraphics[width=0.31\columnwidth]{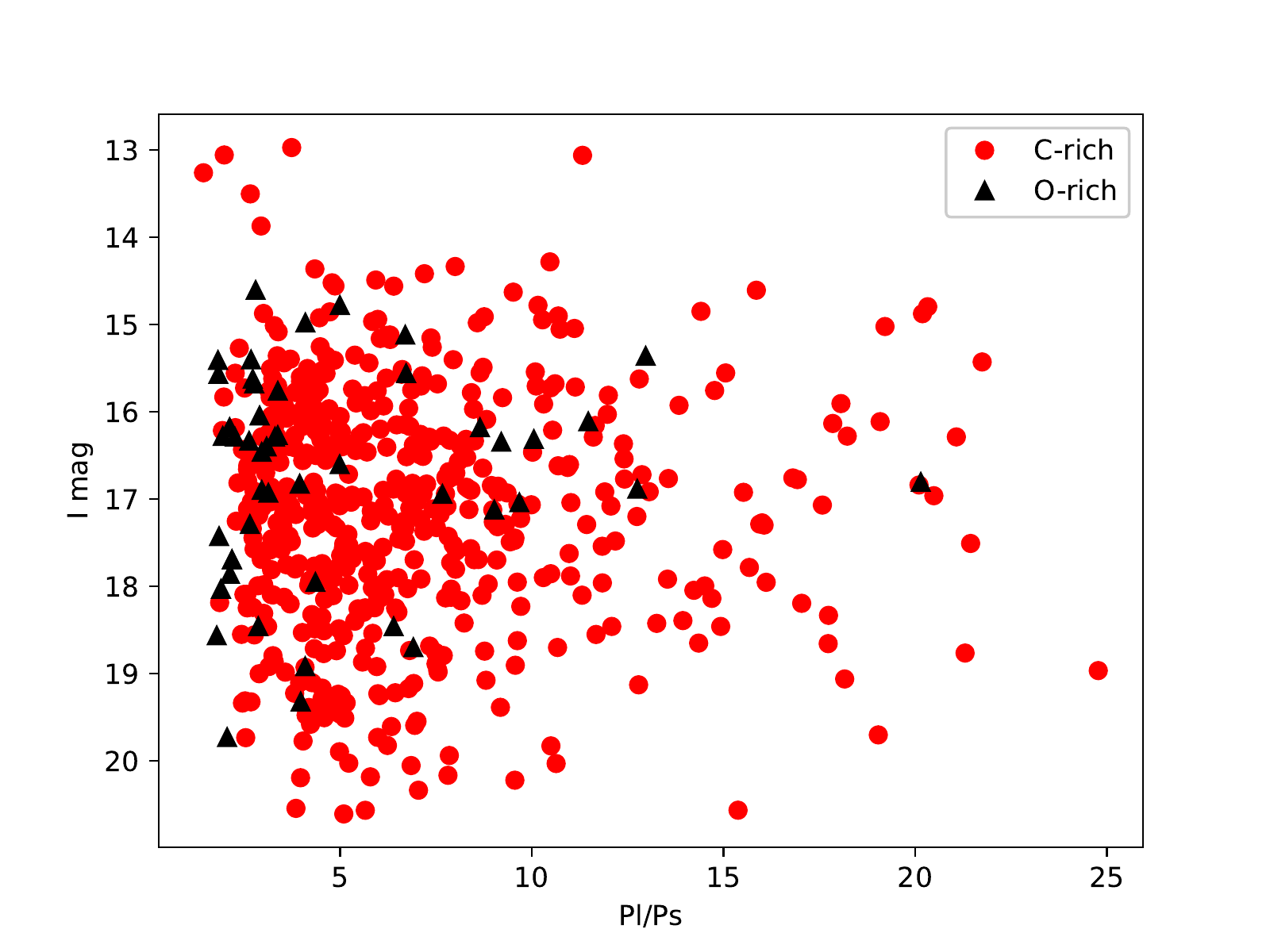} 
    \end{tabular}
    \caption{The PL relations for the non-regular Miras in the LMC, using either the $P_s$ (left panel), the $P_l$ (middle panel), or the $P_l/P_s$ ratios (right panel) as independent variable.}

    \label{multi-pl}
\end{figure*}

For the non-regular Miras classified in Section 2, we fitted their OGLE-III $I$-band light curves with two components: 

\begin{eqnarray}
I(t) & = & I_0 + L^s(t) + L^l(t), \\
 & = & I_0 + \sum_{i=1}^3 A^s_i \cos (2i\pi\frac{t}{P_s} + \phi^s_i) +  \sum_{i=1}^3 A^l_i \cos (2i\pi\frac{t}{P_l} + \phi^l_i).
\end{eqnarray}

\begin{figure}
    \centering
    \plottwo{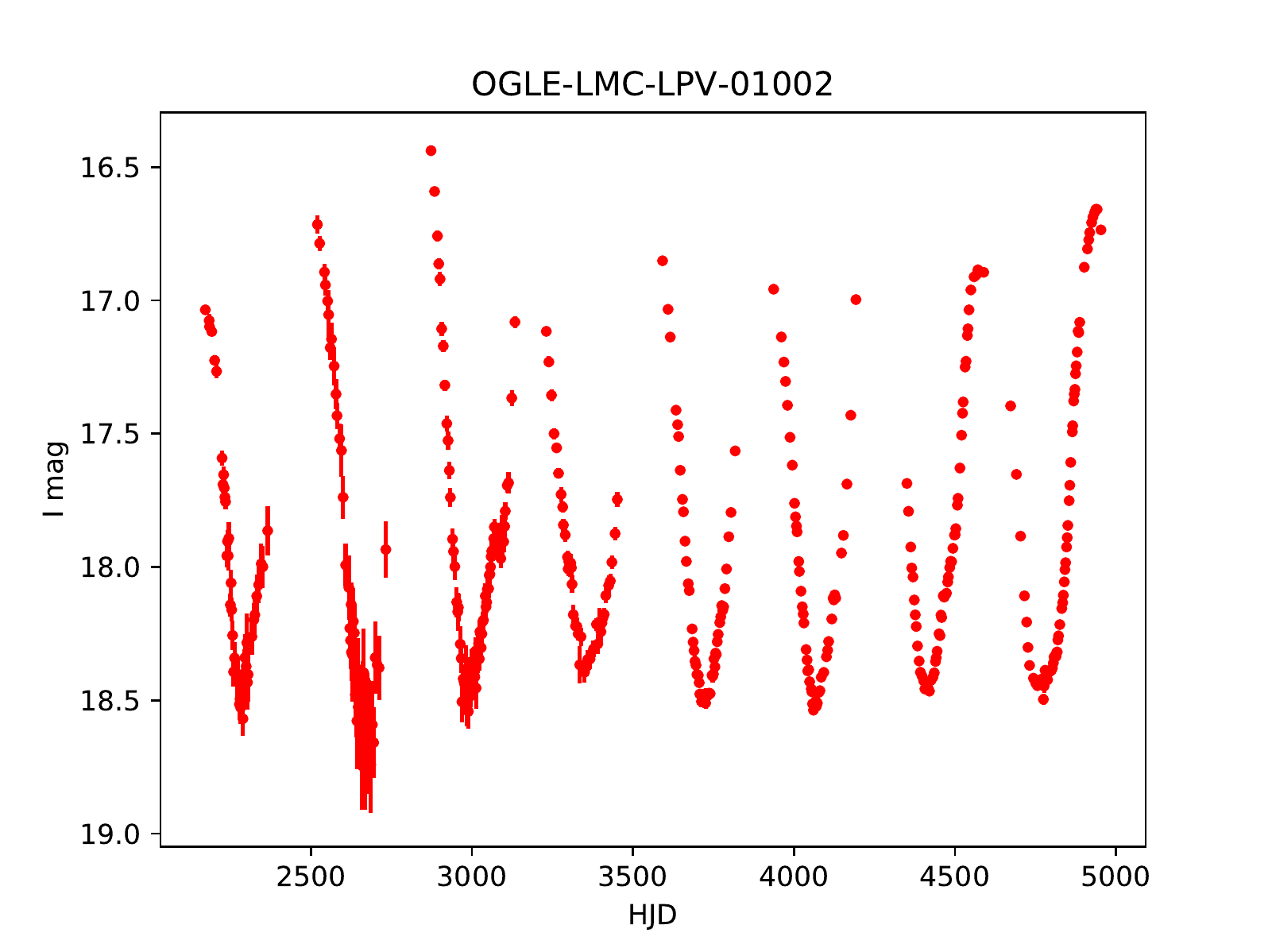}{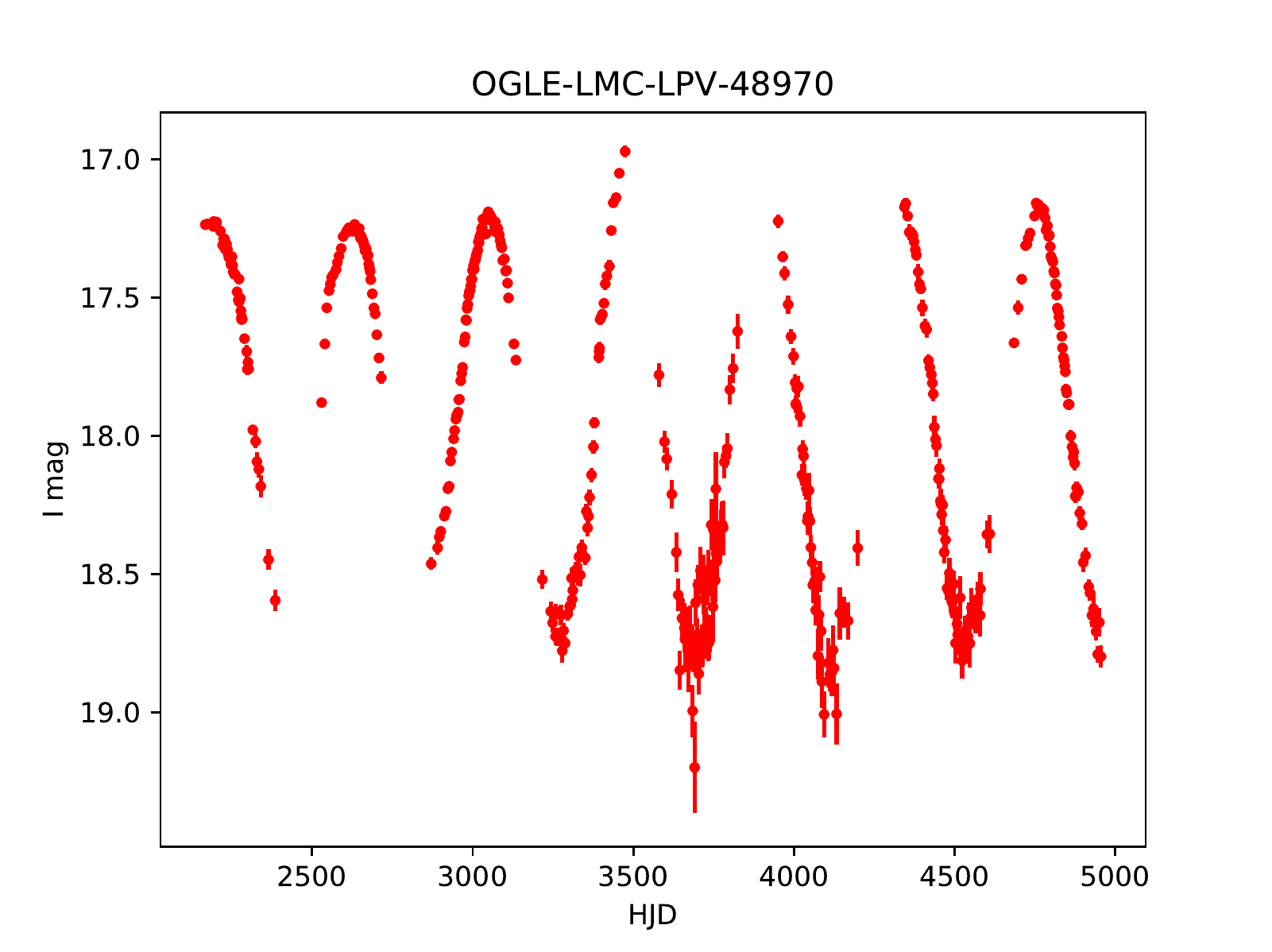}
    \plottwo{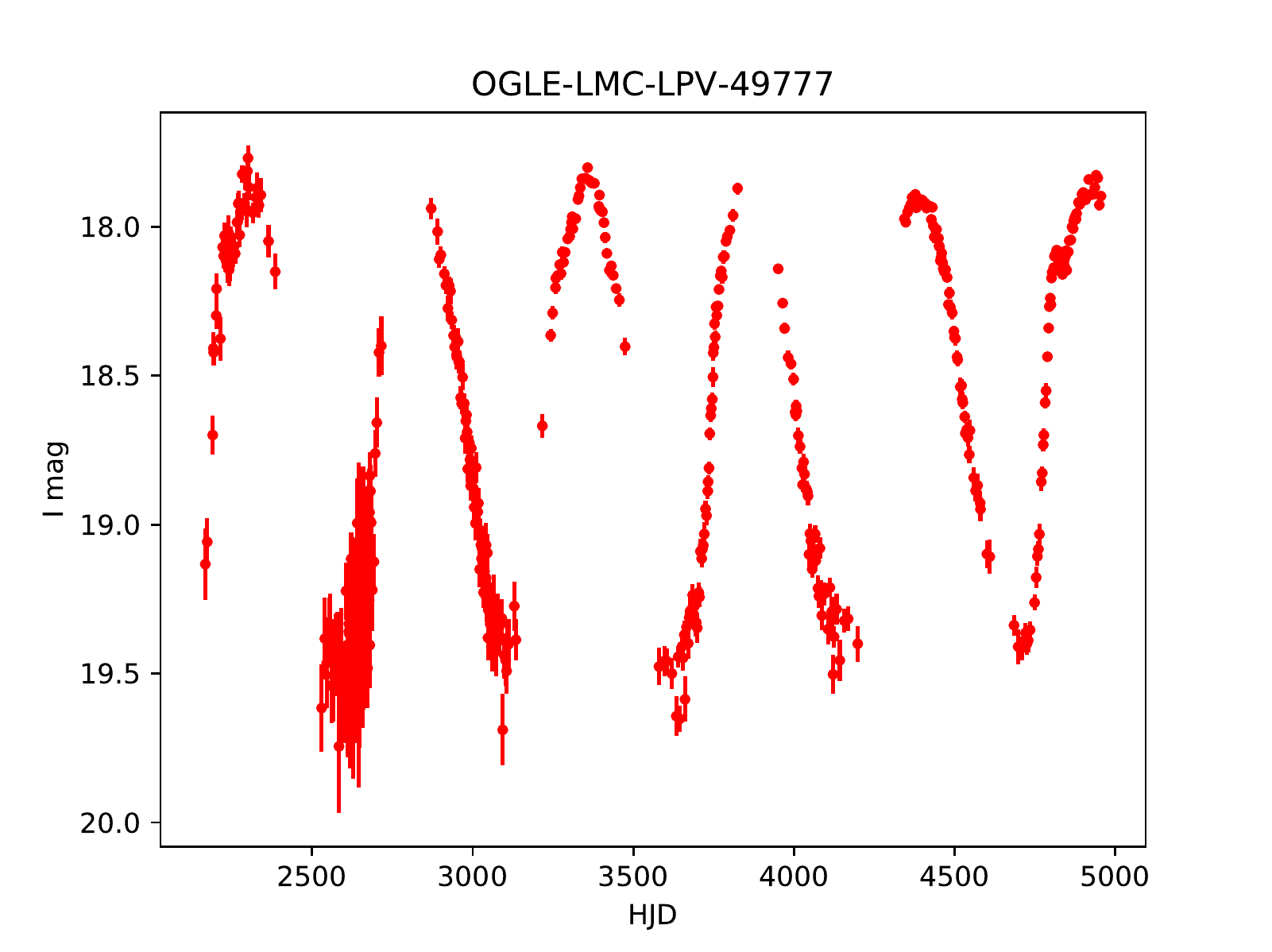}{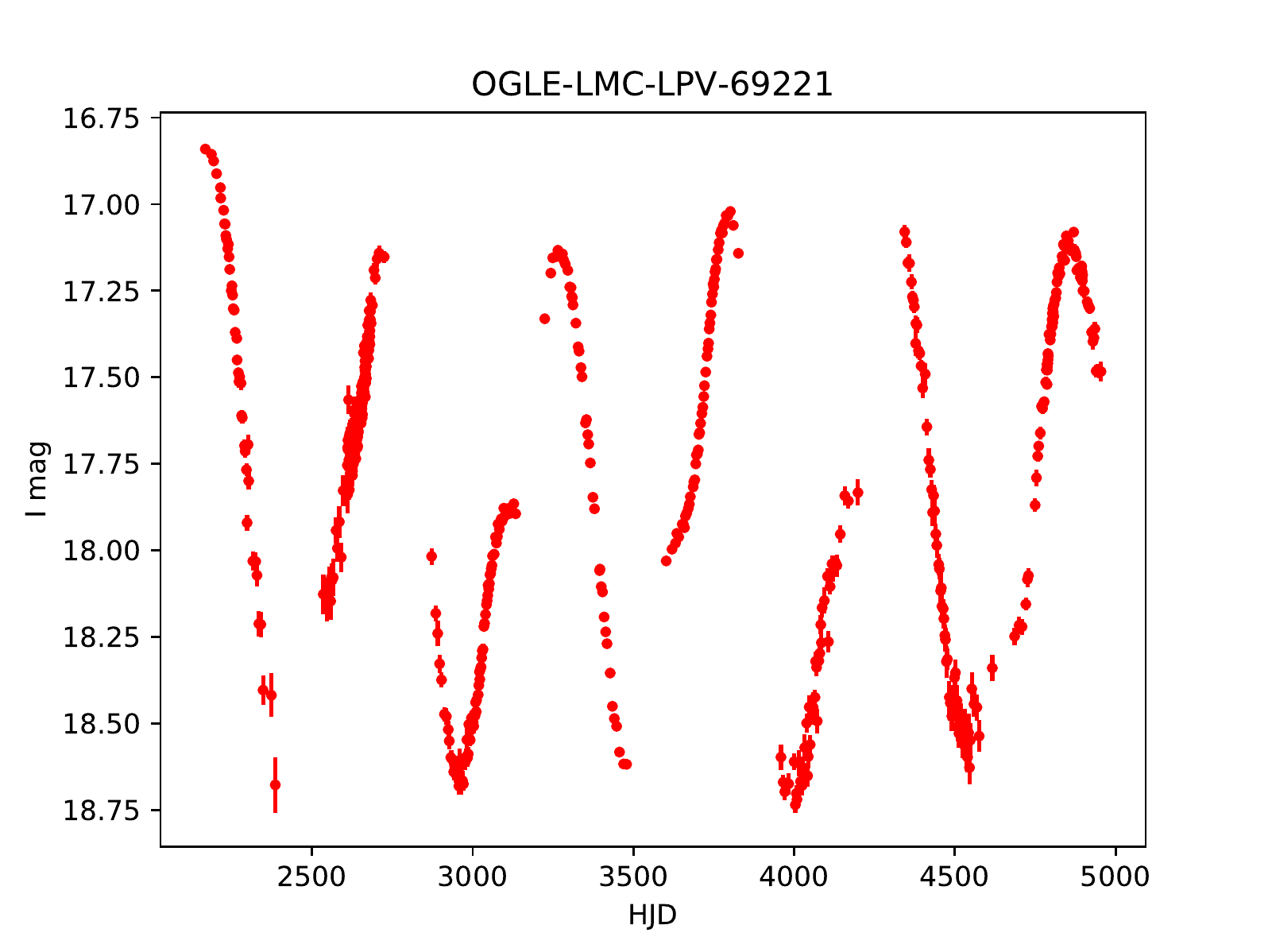}
    \caption{Examples of light curves for non-regular Miras after removing the long-term variation $L^l(t)$.}
    \label{rml}
\end{figure}

\noindent In the above equations, $P_s$ represents the short-term or primary pulsation period, and $P_l$ is the long-term period. We applied the Nelder–Mead algorithm to search for the best-fit $P_s$-$P_l$ pairs by evaluating the $\chi^2$ values from the fitted light curves. We restricted the range of $P_s$ to be between 100 to 1000~days  (using the periods found from the LS periodogram, $P_{LS}$, as initial guesses) and $P_l$ to be between 1000 to 5000~days. To speed up the calculations, a pair of $P_s$-$P_l$ was set as a starting point, then about eight pairs around the starting point were selected to fit the function. The pair with the smallest $\chi^2$ value served as the starting point for the next iteration. This procedure was repeated until a local minimum was reached, and the best fitting values of $P_s$ and $P_l$ were determined, as presented in Table \ref{fitper_data}. We picked 10 non-regular Miras to perform a Monte Carlo simulation by simulating 100 light curves for each of them using the observed errors. These simulated light curves were then run through the same Nelder–Mead algorithm. Based on our simulations, we estimated the $P_l$ error to be about 100~days. The distribution of the $P_l/P_s$ ratio was shown in the left panel of Figure \ref{hist_multi}, which peaks around 5 with a long tail toward a ratio of $\sim20$ for the C-rich non-regular Miras. Furthermore, no correlation was found between the $P_l/P_s$ ratio and the pulsation period $P_s$, as shown in the right panel of Figure \ref{hist_multi}, implying $P_l$ was independent of the pulsational properties of Miras. Figure \ref{lc_multi} presents several examples of the non-regular Miras, together with the best-fit $I(t)$ to their $I$-band light curves based on the determined $P_s$-$P_l$ pairs.


\begin{deluxetable*}{llCCC}
\label{fitper_data}
\tabletypesize{\footnotesize}
\tablecaption{Periods of the non-regular Miras. 
			}
\tablecolumns{5}
\tablewidth{0pt}
\tablehead{
	\colhead{MIRA\_ID} & \colhead{Spectra type} & \colhead{$P_s$ (days)} & \colhead{$P_l$ (days)} &  \colhead{$P_{OGLE}$ (days)}
          }
\startdata
OGLE-LMC-LPV-00055& C-rich & 	288.98 &	6089.38 & 290.9\\
OGLE-LMC-LPV-00094& C-rich & 	331.56 &	2089.78 & 332.3\\
OGLE-LMC-LPV-00144& C-rich & 	370.50 &	2727.32 & 364.2\\
OGLE-LMC-LPV-00225& C-rich & 	494.30 &	5974.18 & 504.2\\
$\cdots$ & $\cdots$ & $\cdots$ & $\cdots$ & $\cdots$ \\
\enddata
\tablecomments{The entire Table is published in its entirety in the machine-readable format. A portion is shown here for guidance regarding its form and content.}
\end{deluxetable*}

\begin{figure*}
\plotone{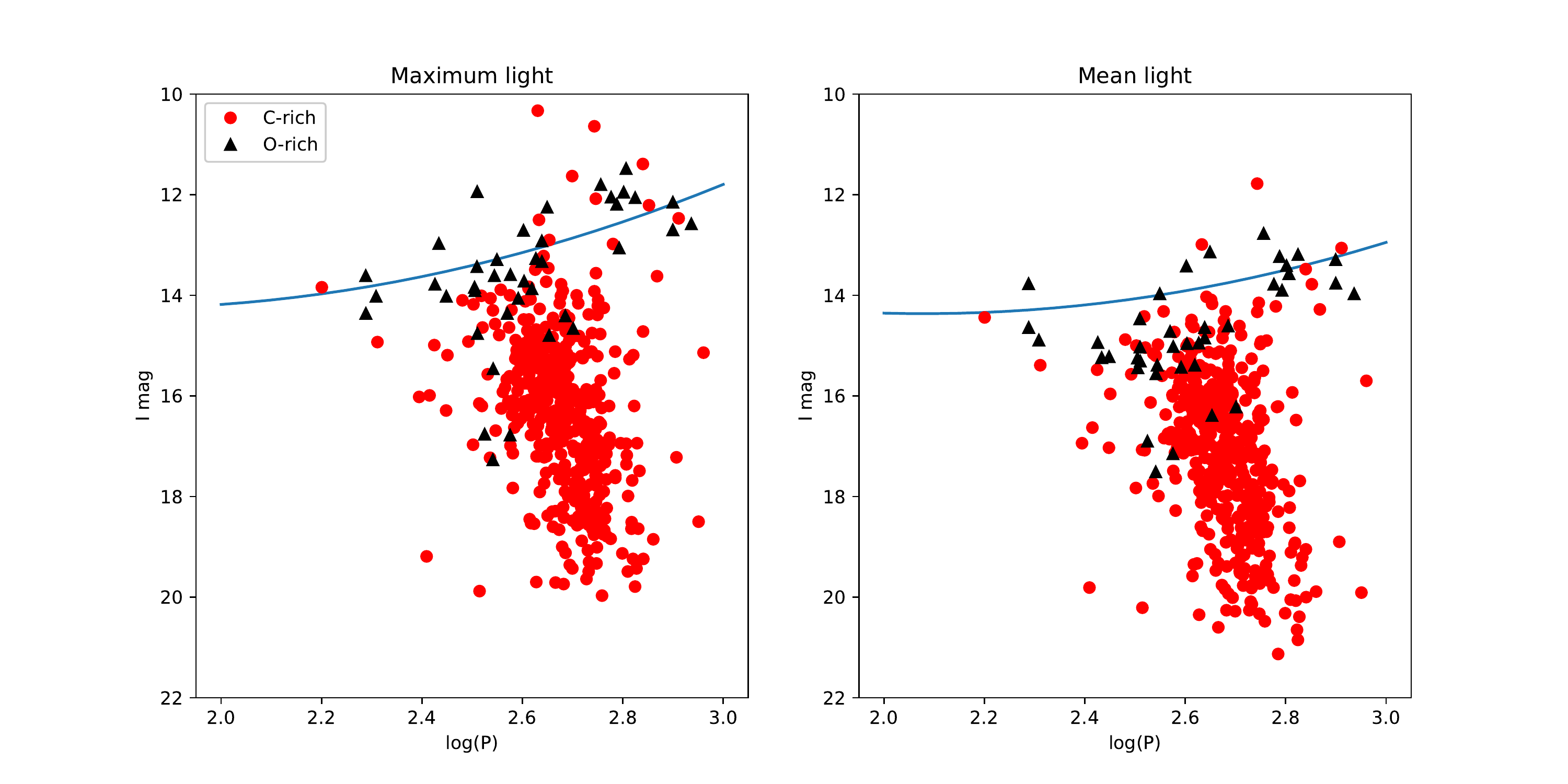}
\caption{The PL relations at maximum light (left panel) and at mean light (right panel) for non-regular Miras after removing the long-term variations. The curves are the fitted quadratic models as given in Table \ref{tab2} and not the fittings to the O-rich non-regular Miras.}
\label{rmpl}
\end{figure*}

It is clear that magnitudes at both the maximum and minimum light are unstable, thus showing non-regular Miras unsuitable for deriving PL relations. This is further confirmed in Figure \ref{multi-pl}, which displays non regular Miras having no clear trend in the PL relations. We have also removed the long-term variations, $L^l(t)$ in Equation 3, and only kept the $I_0 + L^s(t)$ for the non-regular Miras light curves. An example shown in Figure \ref{rml} suggested that the resulting light curves were similar to the regular Miras. Using the same methodology described in Section 3, we derived the PL relations at both maximum and mean light after removing $L^l(t)$. The resulting PL relations, as presented in Figure \ref{rmpl}, do not exhibit a clear correlation similar to regular Miras. We also found that the scatter was larger at maximum light than the mean light for these PL relations. However, when comparing the left panel of Figure \ref{multi-pl} to \ref{rmpl}, the O-rich non-regular Mira, after removing the $L^l(t)$ term, located closer to the quadratic PL relations based on the regular Miras.
In contrast, the C-rich non-regular Miras did not correlate between the pulsation periods and magnitudes at neither mean nor maximum light. Based on the previous studies, C-rich Mira that presented a long-term trend on its light curve may be due to the presence of circumstellar reddening \citep{1984MNRAS.211..331F}, similar to some AGB stars that had a feature of a thick dust shell \citep{2003MNRAS.342...86W}. We speculated dust might play some important roles for the C-rich non-regular Miras. 



\section{SED Fittings}

\begin{figure*}
    \centering
    \plottwo{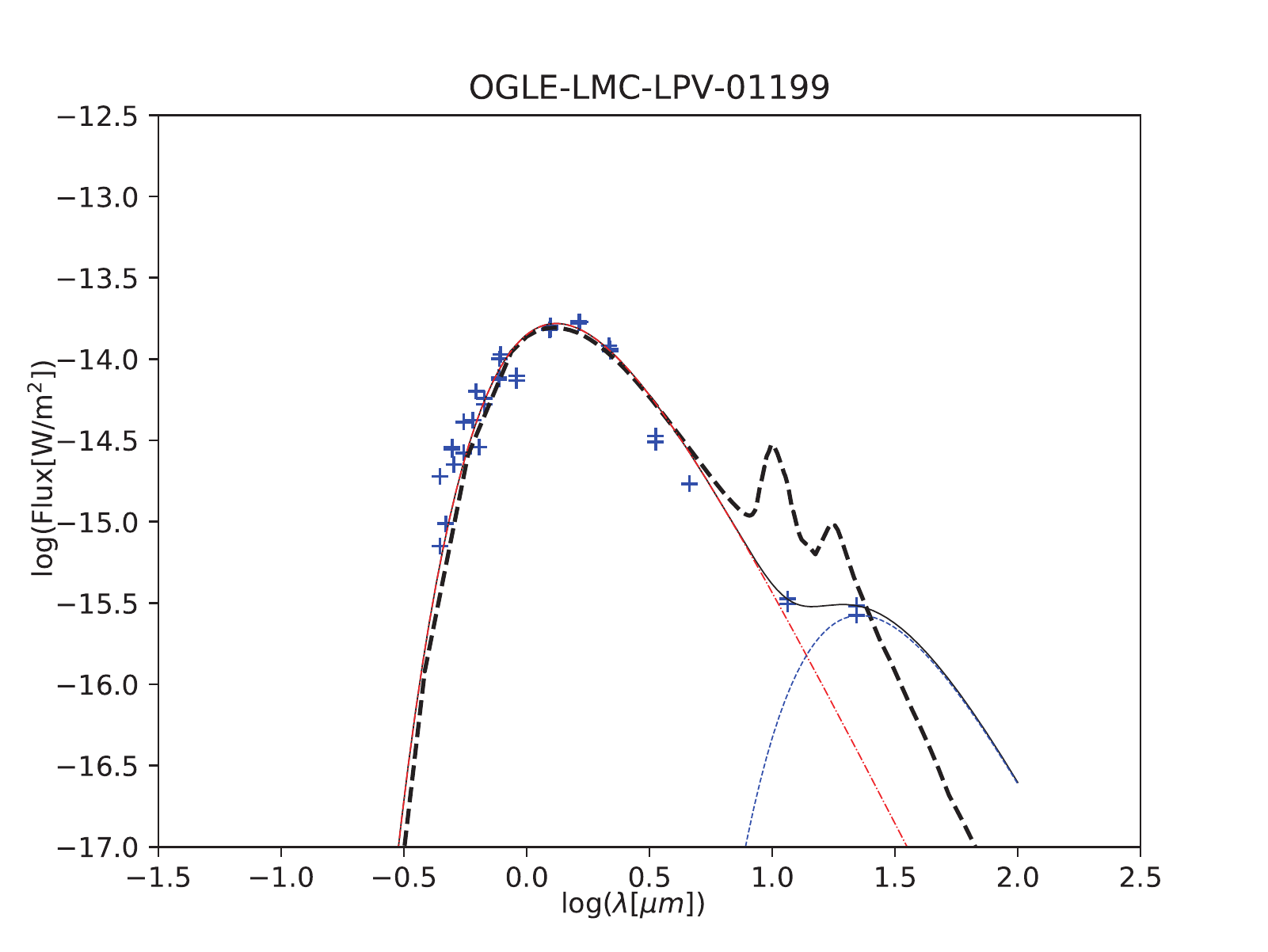}{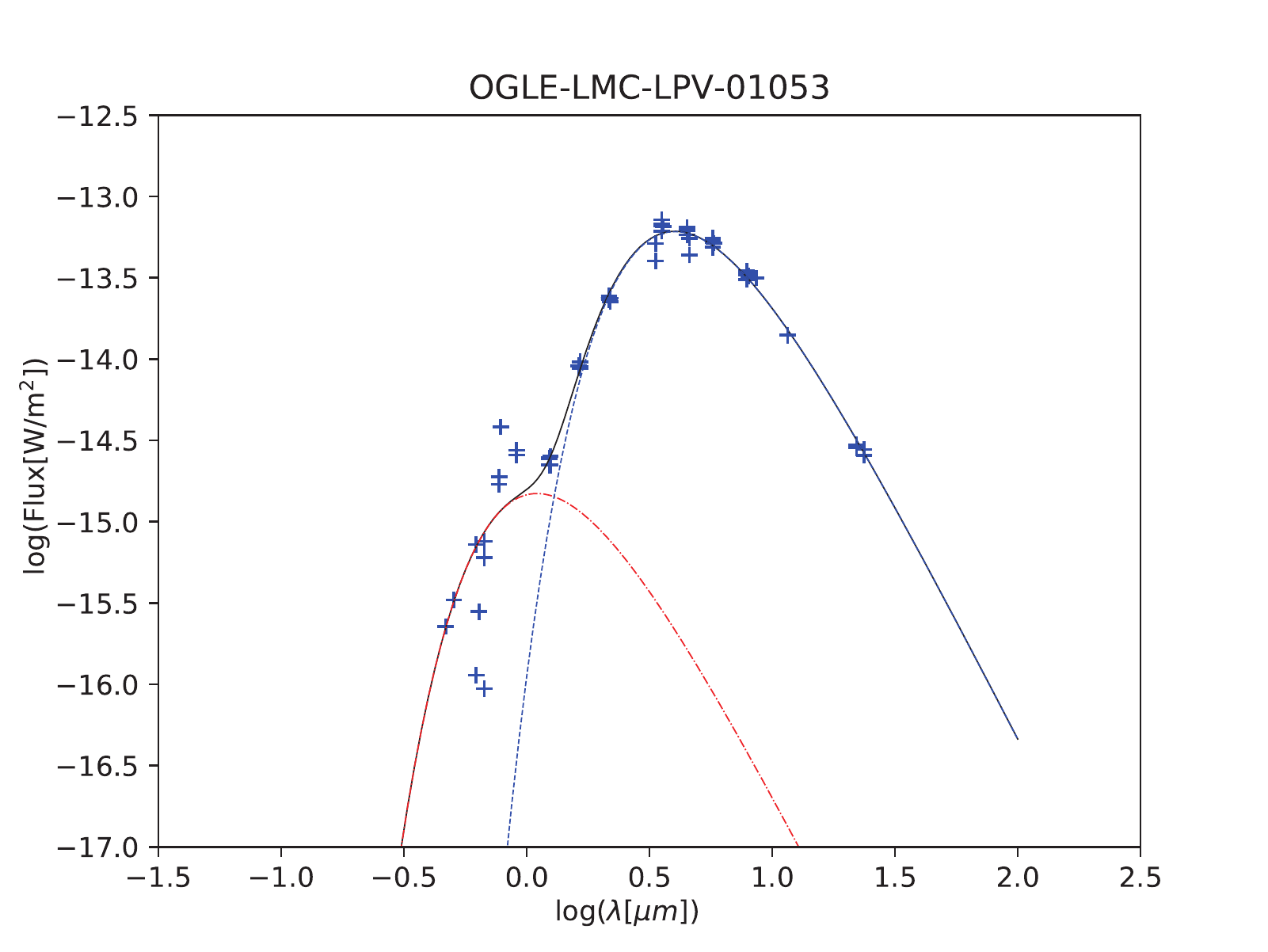}
    \plottwo{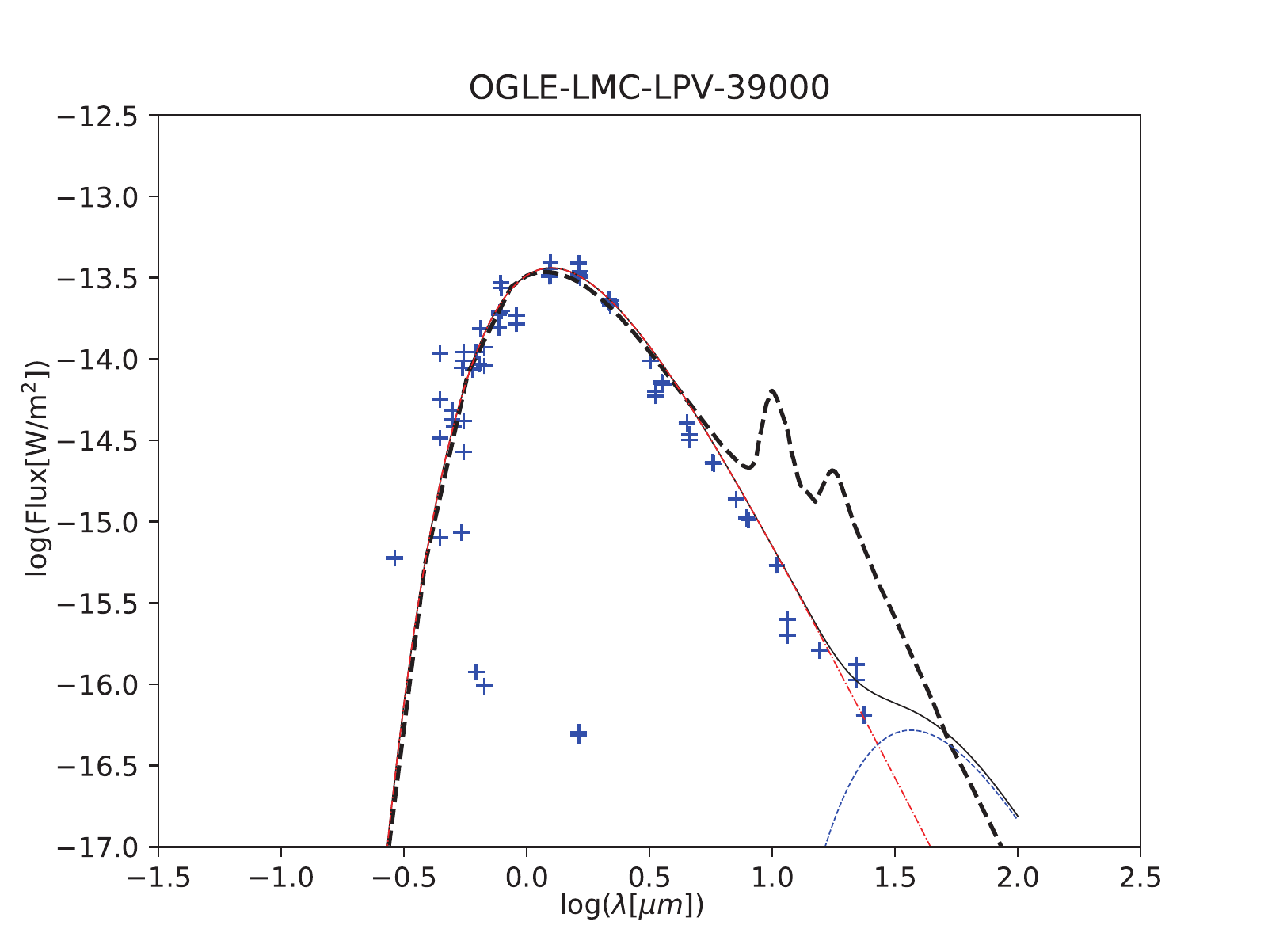}{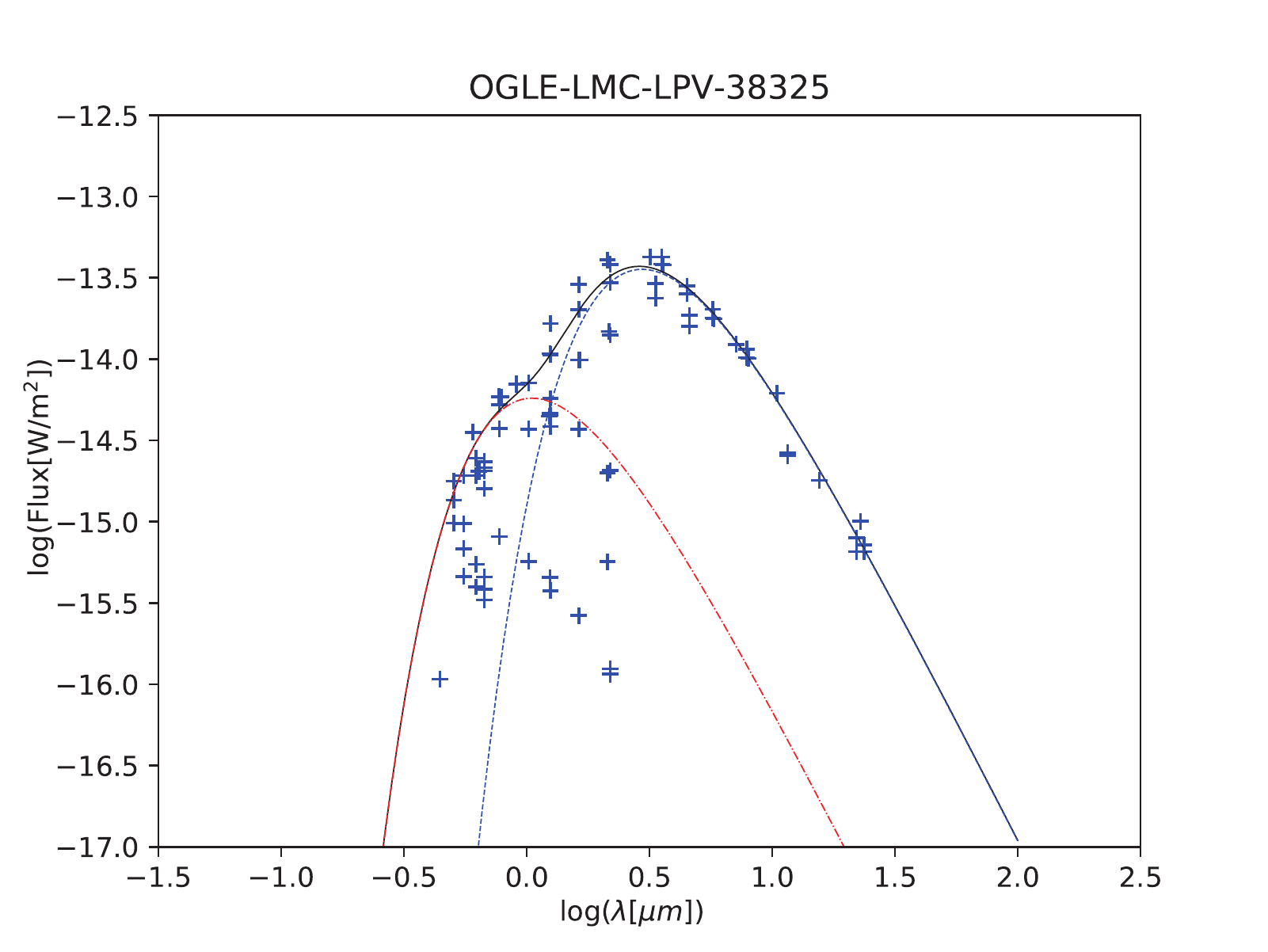}
    \plottwo{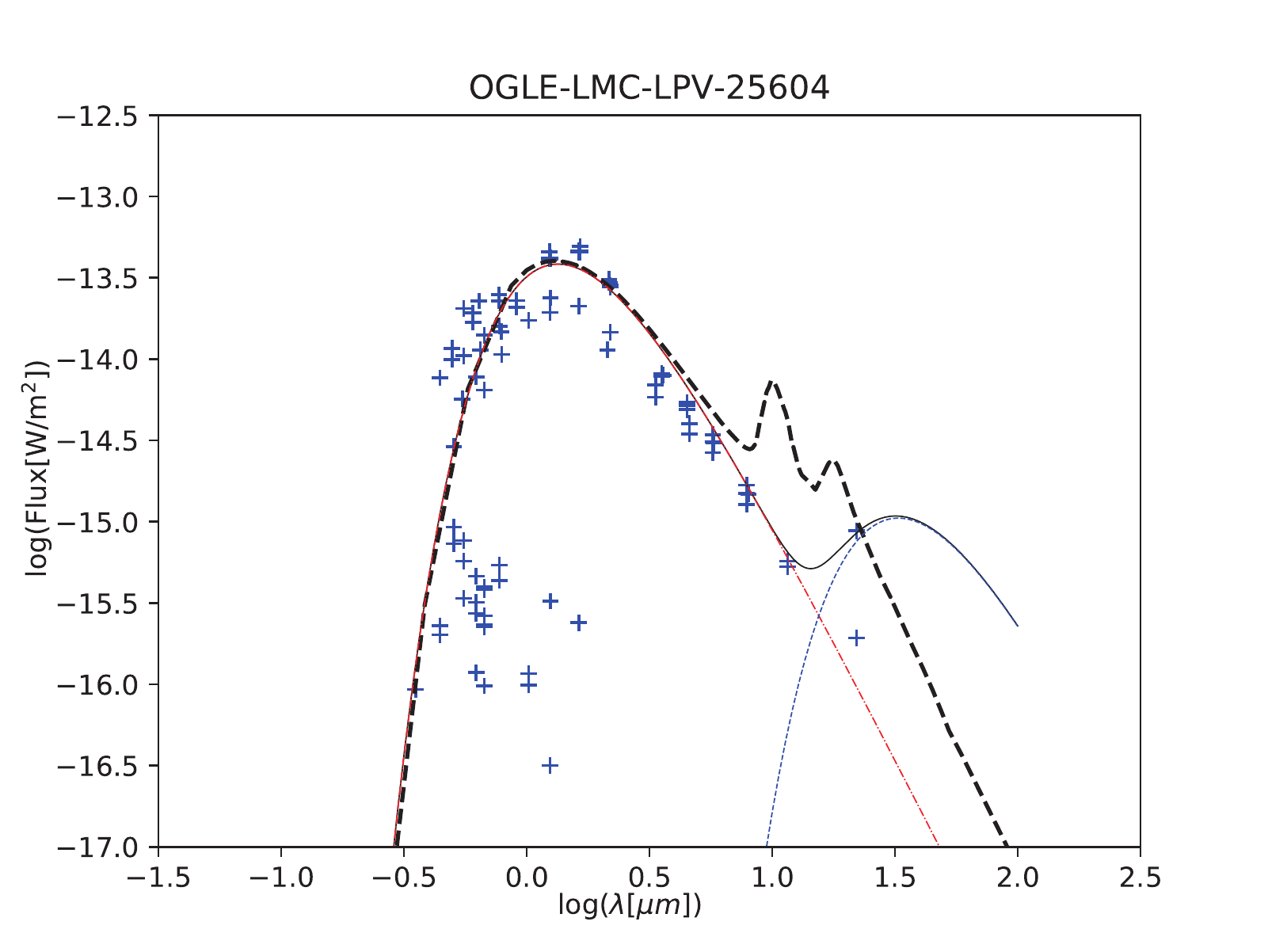}{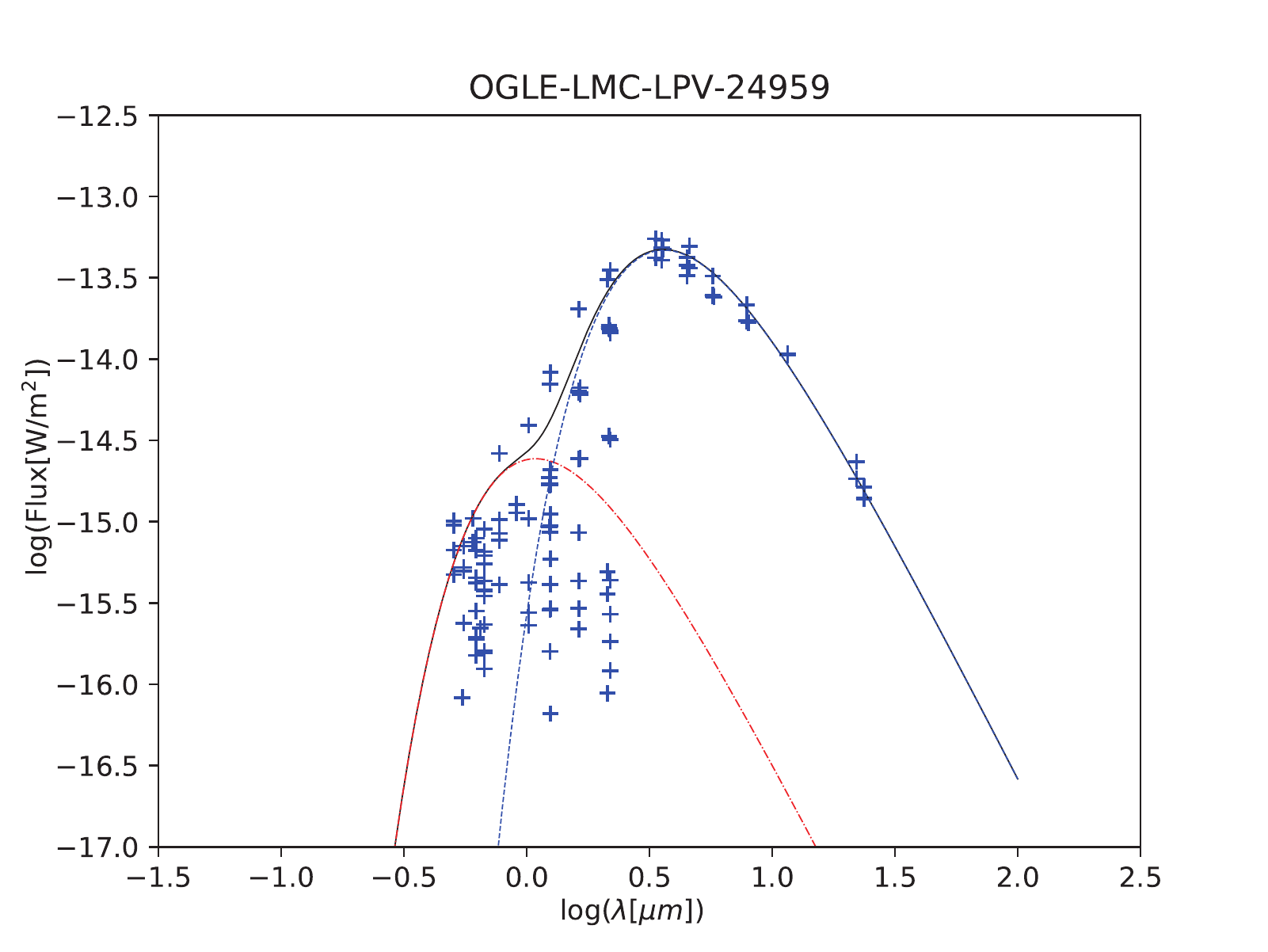}
    \caption{Examples of SED fittings for regular Miras (left column) and non-regular Miras (right column). The data points in blue crosses are the available photometric data taken from the SIMBAD database. The red dashed and blue dotted curves are the best-fit black-body radiation function for representing the star and the dust components, respectively. The sum of these two components is shown as the black solid curves. In the left column, the black dashed curves are the resulting SED fit using the DESK model (see text for details).  }
    \label{SED}
\end{figure*}

\begin{figure}
    \centering
    \plottwo{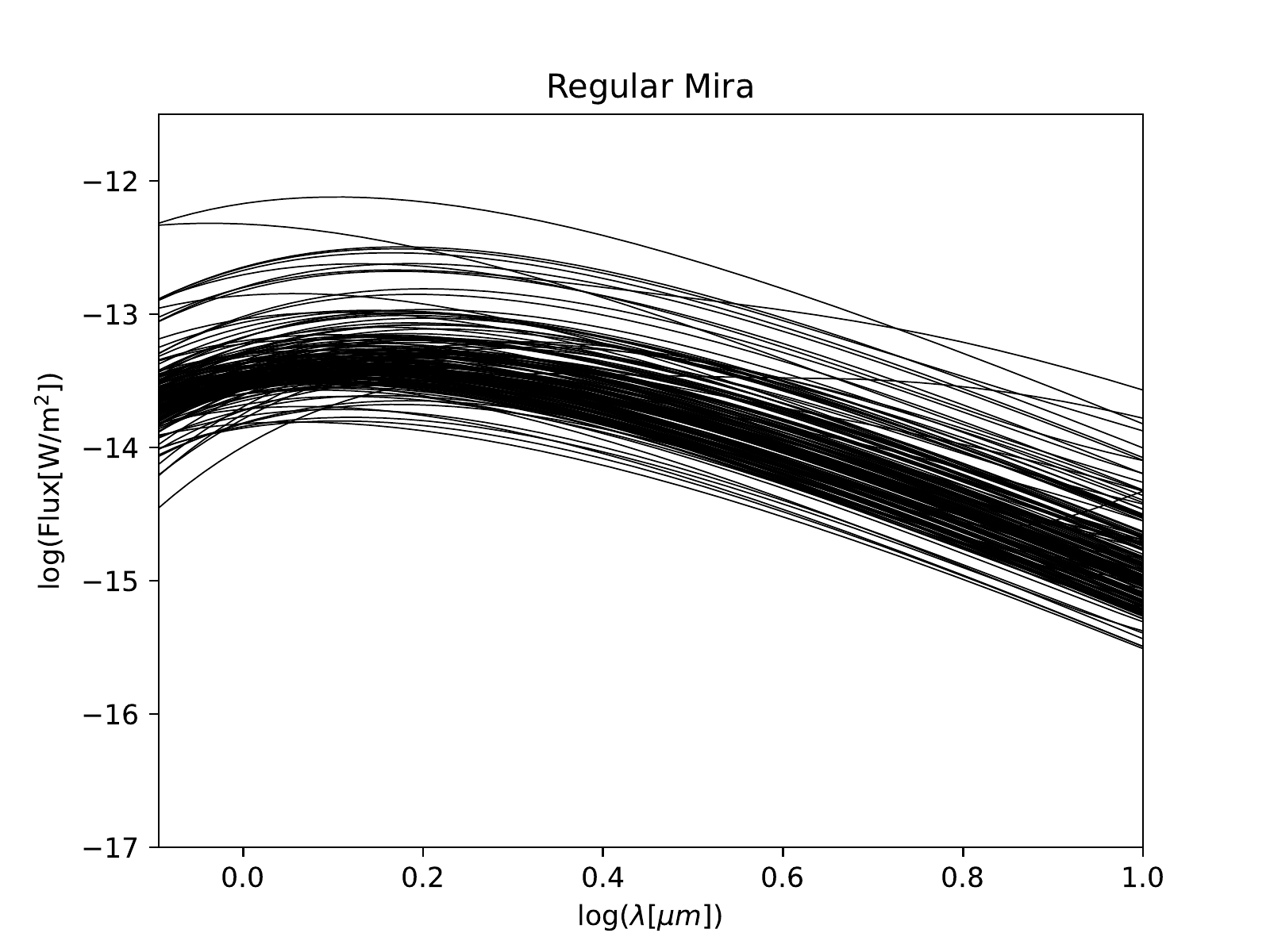}{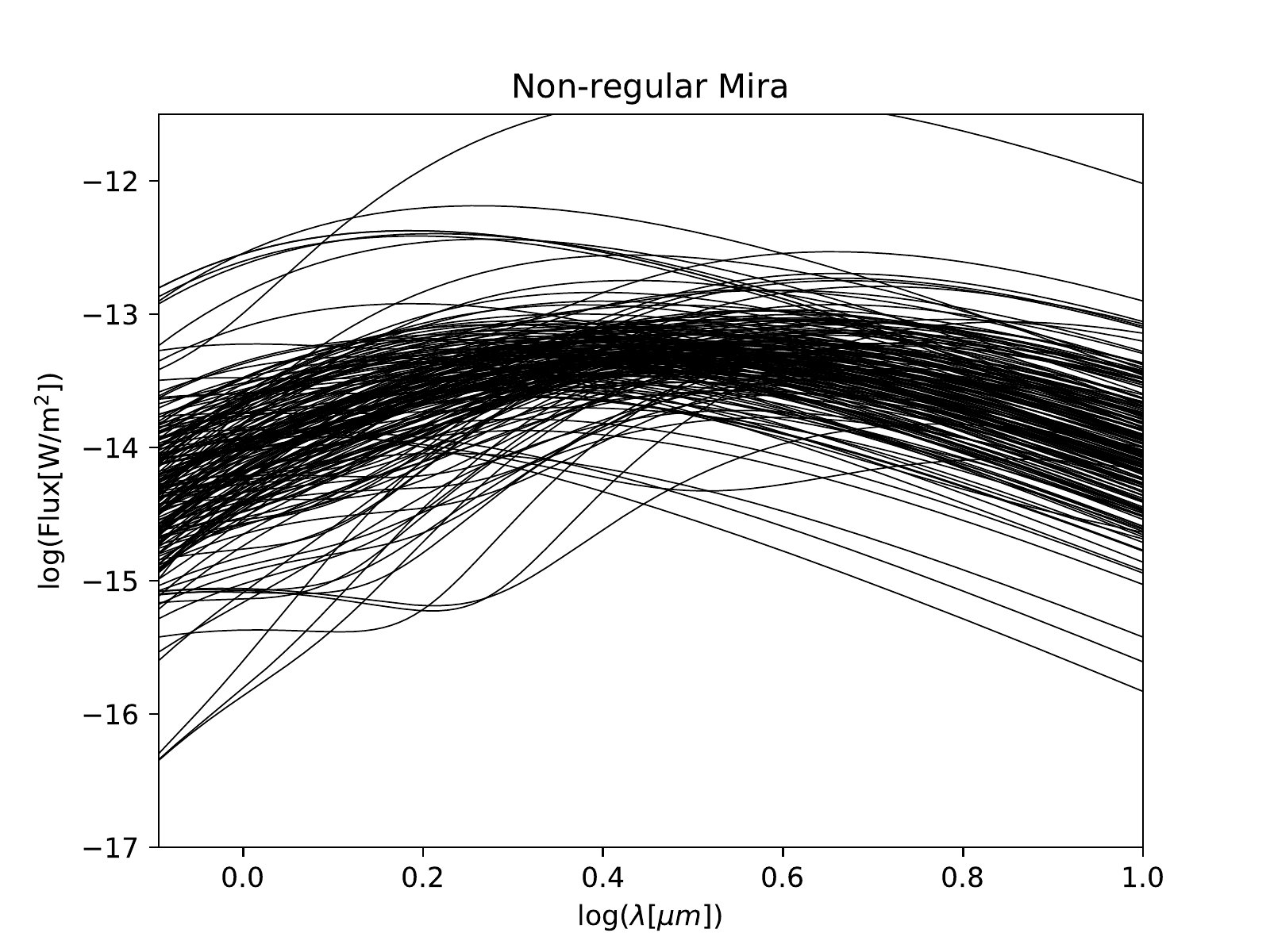}
    \caption{ Composite of the fitted SED between the $I$-band and 10~micron. The left for the regular Miras and the right for non-regular Miras. It is clear that the peaks of the SED for regular Miras are around $\sim1.26$~micron (or $\log \lambda \sim 0.1$), but for non-regular Miras the peaks are around $\sim1.6$ to $\sim6.3$~micron (or $\log \lambda \sim 0.2$ to $\sim0.8$). }
    \label{SEDs_I10}
\end{figure}

For our sample of Magellanic Clouds Miras, we have also collected their multi-band photometry from the SIMBAD database, whenever available, to perform the spectral-energy-distribution (SED) fittings  \citep[at which the same SIMBAD database and methodology was applied in][]{2021ApJ...911...51L}. This multi-band photometry was mostly single-epoch random-phase measurements from various catalogs (see Table \ref{SED source}). Nevertheless, we fitted our SED only to the maximum values for a given filter. These included 196/143 and 28/56 regular/non-regular Miras in the LMC and SMC. We pointed out that for the case of non-regular Miras, the multi-band photometry from the SIMBAD database can only be found for the C-rich Miras to perform the SED fitting. Extinctions for each filter, $A_{\lambda}$, were corrected using the optical extinction map from \citet{2021ApJS..252...23S} together with the reddening law adopted from \citet{1989ApJ...345..245C}. Our model for the SED fitting consists of two black-body radiation functions to represent the star and the dust components. These two components were modeled as projected 2-D "disk" from two spheres with different radii. When performing the SED fitting, we restricted the ranges of the black-body temperature between 2000~K and 4000~K for the stellar component and between 10~K to 1800~K for the dust component. Same as in Section 3, we adopted a distance of 49.59~kpc and 62.44~kpc, for LMC and SMC, respectively, when converting the (extinction-corrected) multi-band observed magnitudes to the fluxes. Comparisons of the fitted SED for both Miras showed that the peak of the fitted black-body curve for the dust component in non-regular Miras was much higher than the stellar component, as presented in Figure \ref{SED} for six examples. This suggested that non-regular Miras had more dust than regular Miras. Figure \ref{SEDs_I10} shows the composite fitted SEDs at infrared for the regular and non-regular Miras.


\begin{deluxetable}{ccc}
\label{SED source}
\tabletypesize{\footnotesize}
\tablecaption{ List of broadband photometric data catalogs used in SED fitting.}
\tablecolumns{3}
\tablewidth{0pt}
\tablehead{
	\colhead{Catalog Name} & \colhead{Filter} & \colhead{References}
          }
\startdata
NOMAD-1 Catalog&Ks,H,J&\citet{2004AAS...205.4815Z}\\
The Guide Star Catalog Version 2.3.2&i,F,J &\citet{2008AJ....136..735L}\\
The PPMXL Catalog& Ks,H,J &\citet{2010AJ....139.2440R}\\
XPM Catalog& Ks,H,J &\citet{2011MNRAS.416..403F}\\
SPM 4.0 Catalog& Ks,H,J,V,B &\citet{2011AJ....142...15G}\\
UCAC4 Catalogue& Ks,H,J &\citet{2012yCat.1322....0Z}\\
APOP& Ks,H,J&\citet{2015AJ....150..137Q}\\
UCAC5 Catalogue&G&\citet{2017yCat.1340....0Z}\\
2MASS All-Sky Catalog of Point Sources&K,H,J &\citet{2003yCat}\\
Gaia DR1&G&\citet{2016AA...595A...1G}\\
Gaia DR2&Grp,G,Gbp&\citet{2018AA...616A...1G}\\
Gaia EDR3&Grp,G,Gbp&\citet{2020yCat.1350....0G}\\
WISE All-Sky data Release& Ks,H,J,W1,W2,W3,W4&\citet{2012yCat.2311....0C}\\
AllWISE Data Release& Ks,H,J,W1,W2,W3,W4& \citet{2014yCat.2328....0C}\\
SkyMapper Southern Sky Survey. DR1.1& g,r,i,z& \citet{2018PASA...35...10W} \\
The CatWISE2020 catalog&W1,W2&\citet{2021ApJS..253....8M} \\
The band-merged unWISE Catalog&W1,W2&\citet{2019ApJS..240...30S}\\
ASAS-SN catalog&Ks,H,J,G,V,W1,W2,W3,W4 &\citet{2018MNRAS.477.3145J}\\
TESS Input Catalog v8.0 &Ks H,J,V,B,W1,W2,W3,W4,&\citet{2019AJ....158..138S}\\
ATLAS all-sky stellar ref. catalog, ATLAS-REFCAT2 &H,J,G &\citet{2018ApJ...867..105T}\\
The HSOY catalogue&H,J,G& \citet{{2017AA600L4A}}\\
VEXAS DR2 catalogs&Ks,J,Y,W1,W2,W3,W4&\citet{2021AA...651A..69K}\\
\enddata
\end{deluxetable}

We have also used a Python package, the Dusty Evolved Star Kit \citep[{\tt DESK},][]{2020JOSS....5.2554G}, to fit the SED and compare it to our result. We used the {\tt Oss-Orich-bb} model and {\tt Zubko-Crich-bb} model available from {\tt DESK}, and the temperatures scale for the stellar component was set in between 2600~K to 3400~K, while for the dusk component, the temperature range is 600~K to 1200~K. In the case of O-rich regular Miras, results from the DESK fitting were consistent with our simple black-body SED fittings. In contrast, DESK predicted a large far-infrared flux excess for the C-rich regular Miras, implying a larger dust abundance than our black-body SED fittings. In the case of non-regular Miras, neither the C-rich nor the O-rich model in DESK can be fitted well to the observed SED. Therefore, we retain a two-components black-body model when fitting the SED to the non-regular Miras.

\begin{figure}
    \centering
    \includegraphics[width=1\columnwidth]{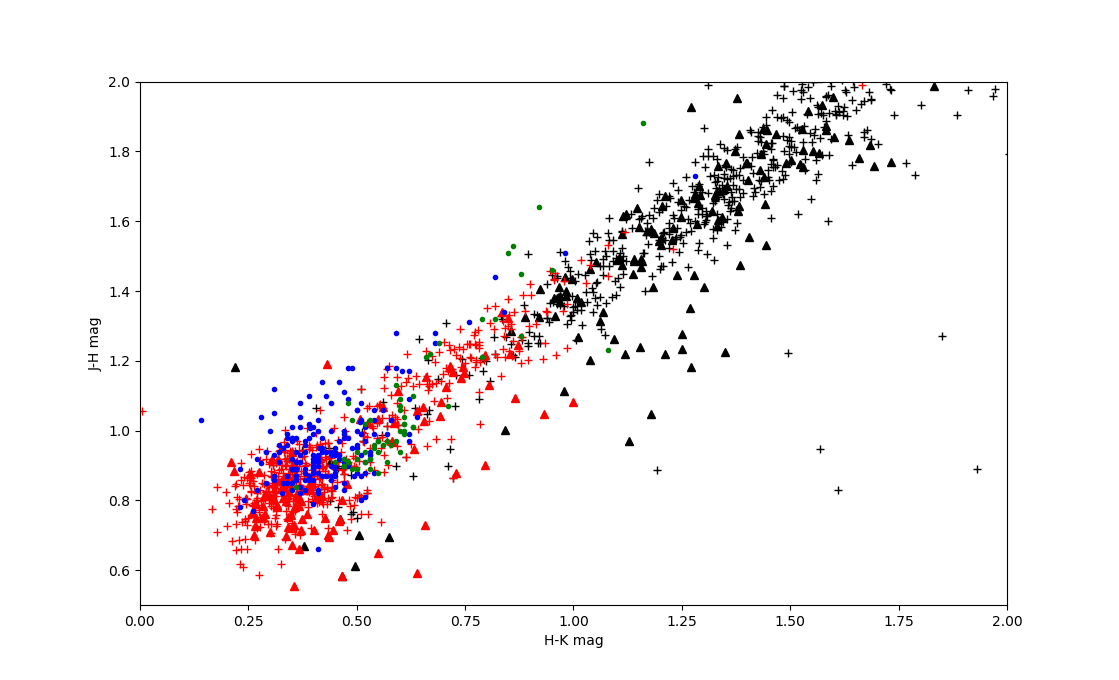}
    \caption{The $JHK$-band color-color diagram for various samples of Miras. The blue and green points are the dust-poor and dust-rich Miras presented in \citet{2000MNRAS.319..728W} and \citet{2000MNRAS.319..759W}, respectively. The red and black crosses are the regular and non-regular Miras in LMC, respectively, based on the single-epoch 2MASS data. Similarly, the red and black triangles are the regular and non-regular Miras in LMC, respectively, based on the multi-epoch LMCNISS data.}
    \label{JHHK}
\end{figure}

Based on the results, the majority of the fluxes for the regular Miras were from the stellar components. In contrast, the dust components for the non-regular Miras contributed most of the overall fluxes. Due to the different levels of infrared excess, dust-rich and dust-poor Miras will have different distributions on the NIR color-color diagram. In Figure \ref{JHHK}, NIR colors in the $JHK$-band for several samples of Miras were compared, including the JHK-band photometry from the Two Micron All Sky Survey (2MASS, \citealt{2003yCat}) and the Large Magellanic Cloud Near-Infrared Synoptic Survey (LMCNISS, \citealt{2017AJ.154..149Y}) for Miras in the LMC. In general, non-regular Miras tend to have larger values of NIR colors (black points in Figure \ref{JHHK}) than that of regular Miras (red points in Figure \ref{JHHK}), even though there were some non-regular Miras having smaller $JHK$-band colors and located within the region occupied by regular Miras (and vice versa). This behavior was similar to the independent samples of dust-rich and dust-poor Miras presented in the literature.

\begin{figure*}
    \centering
    \plottwo{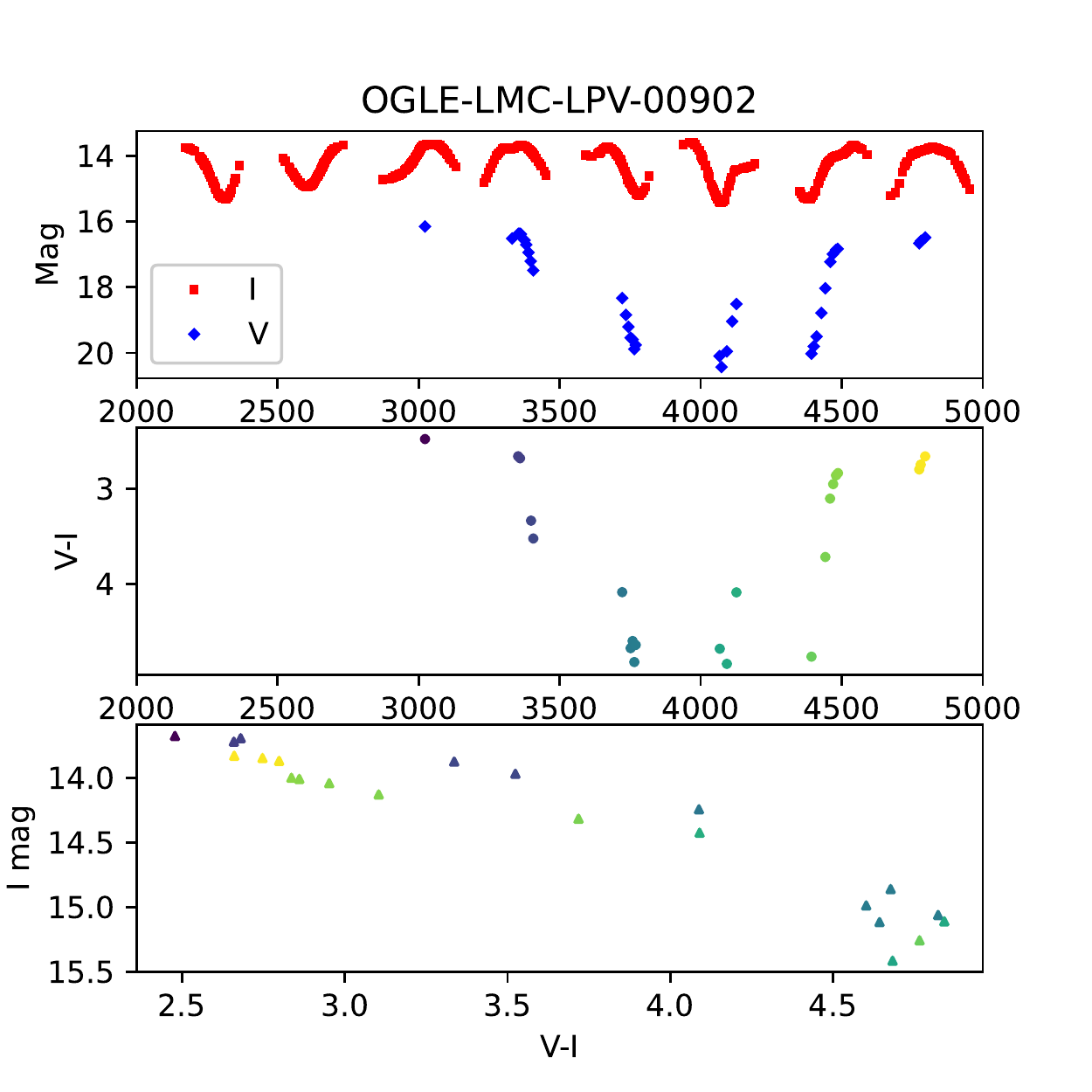}{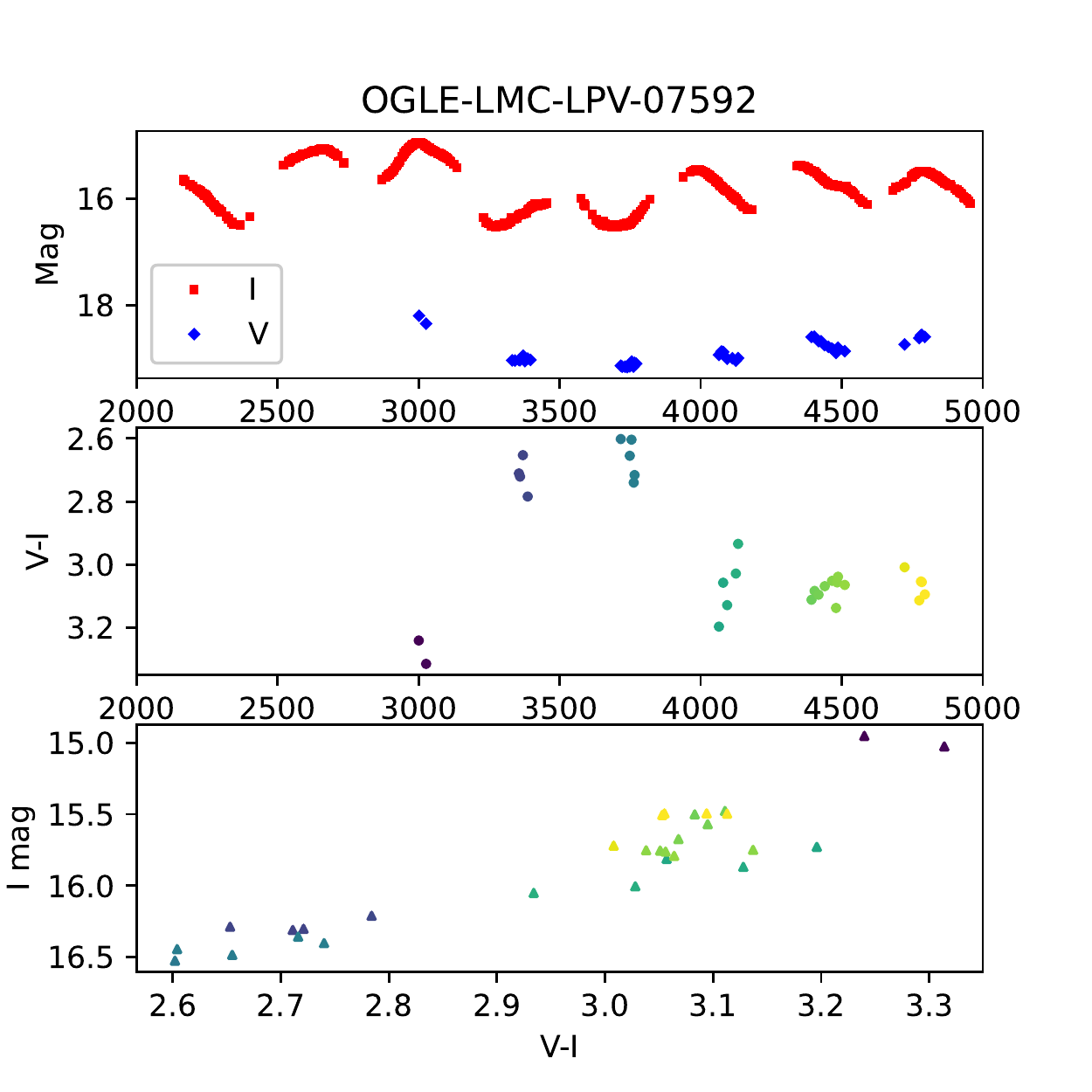}
    \caption{The $VI$-band light curves (top panels), the $(V-I)$ color curves (middle panels) and the corresponding CMD (bottom panels) for a typical regular and non-regular Mira, presented on the left and right panel, respectively. The color of the data points in the bottom two panels corresponds to the time of observations.}  

    \label{soc}
\end{figure*}

\begin{figure*}
    \centering
    \includegraphics[width=1\columnwidth]{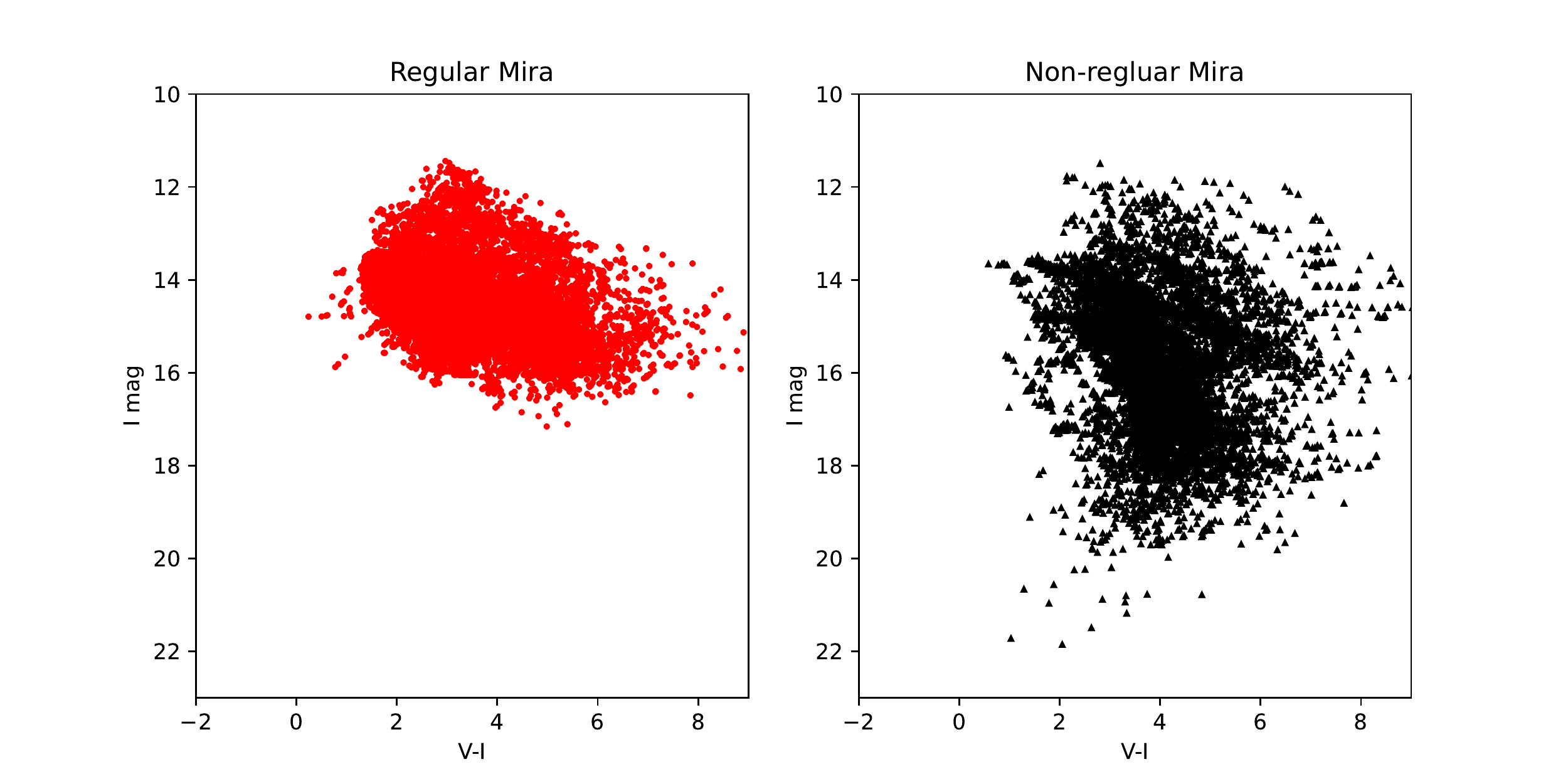}
    \caption{The $VI$-band CMD for the regular Miras (left panel) and non-regular Miras (right panels) in the LMC. Each data-points represents the time-resolved $I$-band magnitudes and the $(V-I)$ colors, based on the OGLE-III light curves, for every available LMC Miras in the OGLE database. 
    }
    \label{oc}
\end{figure*}

The general dust-poor and dust-rich nature of regular and non-regular Miras, respectively, can also be seen from their distinct behaviors in the color-magnitude diagram (CMD). Figure \ref{soc} displays typical examples of the OGLE-III $V,I$-band light curves (top panels), the $(V-I)$ color curves (middle panels, constructed using the $V$- and $I$-band data points that were separated within a day), and the corresponding CMD (bottom panels) at various pulsation phases throughout the pulsation cycles for a regular (left panel) and a non-regular (right panel) LMC Miras. Distributions of the $I$-band magnitudes and $(V-I)$ color from all of the available data points on the $VI$-band light curves for the regular and non-regular LMC Miras are displayed in Figure \ref{oc}. Although there are some overlapping regions, the regular and non-regular Miras occupied different regions on the CMD. The $I$-band magnitudes for regular Miras were confined between $\sim11$ and $\sim17$ mag but showed a large scatter extended to $\sim 22$ mag (due to large variation in their light curves) in the case of the non-regular Miras.

As demonstrated in Figure \ref{soc}, the loci on CMD for the regular Miras followed the expectation of a pulsating star: the color became bluer when the star was brighter (due to Stefan-Boltzmann Law). In contrast, the loci of non-regular Miras on the CMD showed an opposite trend: the overall color became bluer when the star was {\it fainter}. We notice this trend only occurs during or around the minimum of the long-term trend (that could cover a few pulsation cycles, as shown in the right panels of Figure \ref{soc} with data points after the observing time of $\sim3200$~days). For pulsation cycles near the maximum of the long term trend, the behavior of the loci on CMD follows the normal pulsation (the two left-most data points in the middle-right panel of Figure \ref{soc}, with observing time at $\sim 3000$~days and $(V-I)$ colors around 3.2 to 3.3).
The phenomenon of colors becoming bluer when the stars are fainter, especially at or near the flux minimum, is known as the "bluing effect". This effect can be seen in some young stars with circumstellar dust \citep[the UXor-type stars, see][]{1990A&A...236..155B,1995AA...302..472G,1999AJ....118.1043H,2019ApJ...871..183H} due to scattering from the obscured dust.  In the case of nearby Miras and semiregular variables, \citet{2004MNRAS.350..365I} showed that, based on interferometric observations, scattering by dust in the inner circumstellar shell is important, and in general, this could make the color index to be blue. To our knowledge, there is little information in the literature about the photometric variation of optical light in such dust-obscured Miras. We believe that this "blueing effect" is vital in the study of dust-obscured Miras. Again, general trends of redder/bluer colors when the regular/non-regular Miras become fainter can be seen from the composite CMD, as shown in Figure \ref{oc}.

\begin{deluxetable*}{llcCCCC}
\label{fit_data}
\tabletypesize{\footnotesize}
\tablecaption{Results of the SEDs fitting for the regular and non-regular Miras}
\tablecolumns{7}
\tablewidth{0pt}
\tablehead{
	\colhead{MIRA\_ID} & \colhead{Spectra type} & \colhead{Type\tablenotemark{a}} & \colhead{Ts (K)} & \colhead{Rs ($R_\odot$)} & \colhead{Td (K)} & \colhead{Rd ($R_\odot$)} 
          }
\startdata
OGLE-LMC-LPV-00082 & O-rich & R & 3147.43 & 147.67 &  91.73 & 19965.99 \\
OGLE-LMC-LPV-00355 & O-rich & R & 3160.09 & 135.89 & 214.57 & 2704.37 \\
OGLE-LMC-LPV-00743 & O-rich & R & 3014.52 & 213.43 &  15.92 & 19962.04 \\
$\cdots$ & $\cdots$ & $\cdots$ & $\cdots$ & $\cdots$ & $\cdots$ & $\cdots$ \\
\enddata
\tablenotetext{a}{R is for regular Miras, and N is for non-regular Miras}
\tablecomments{The entire Table is published in its entirety in the machine-readable format. A portion is shown here for guidance regarding its form and content.}
\end{deluxetable*}

\begin{figure*}
    \centering
    \plottwo{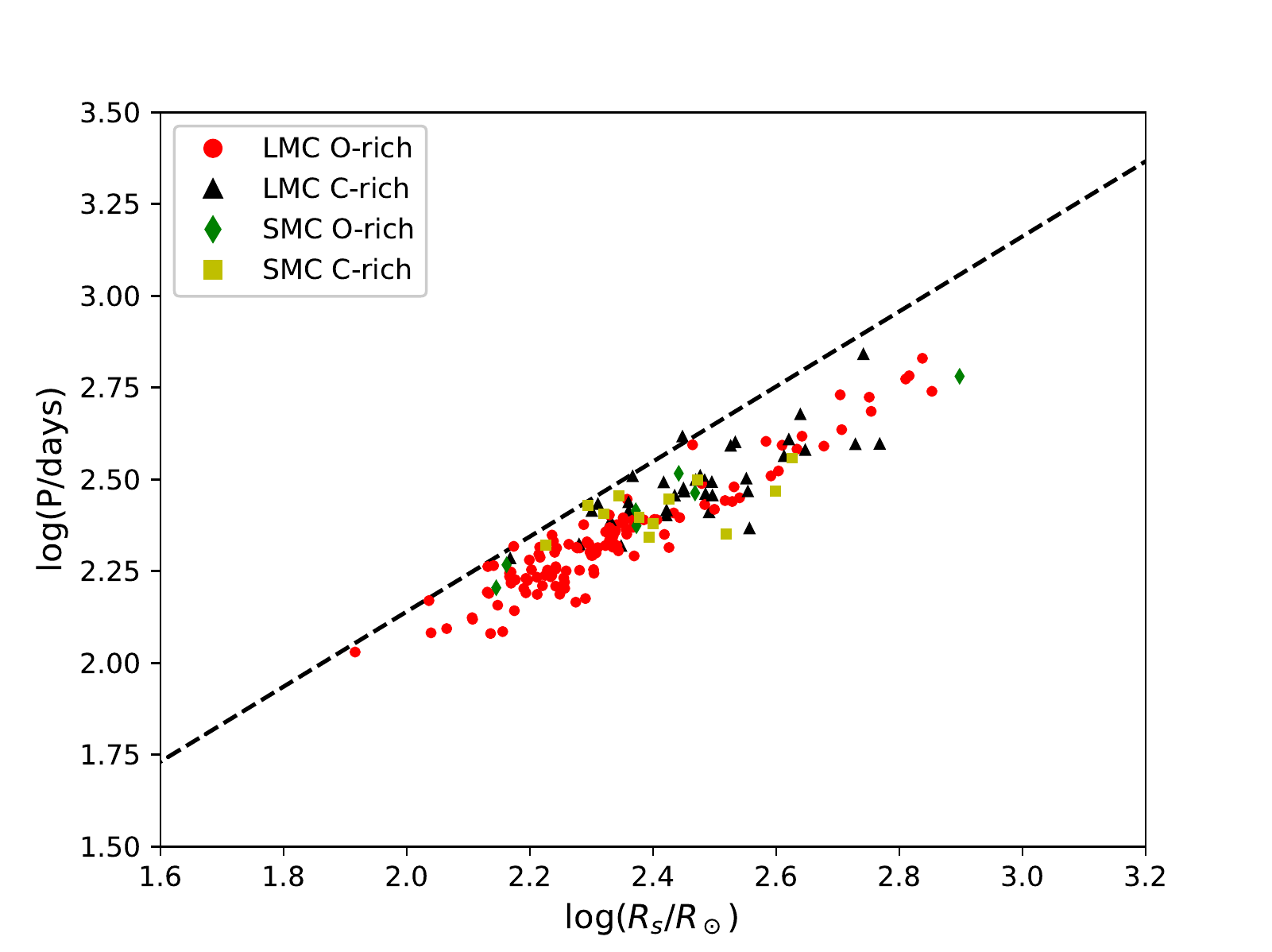}{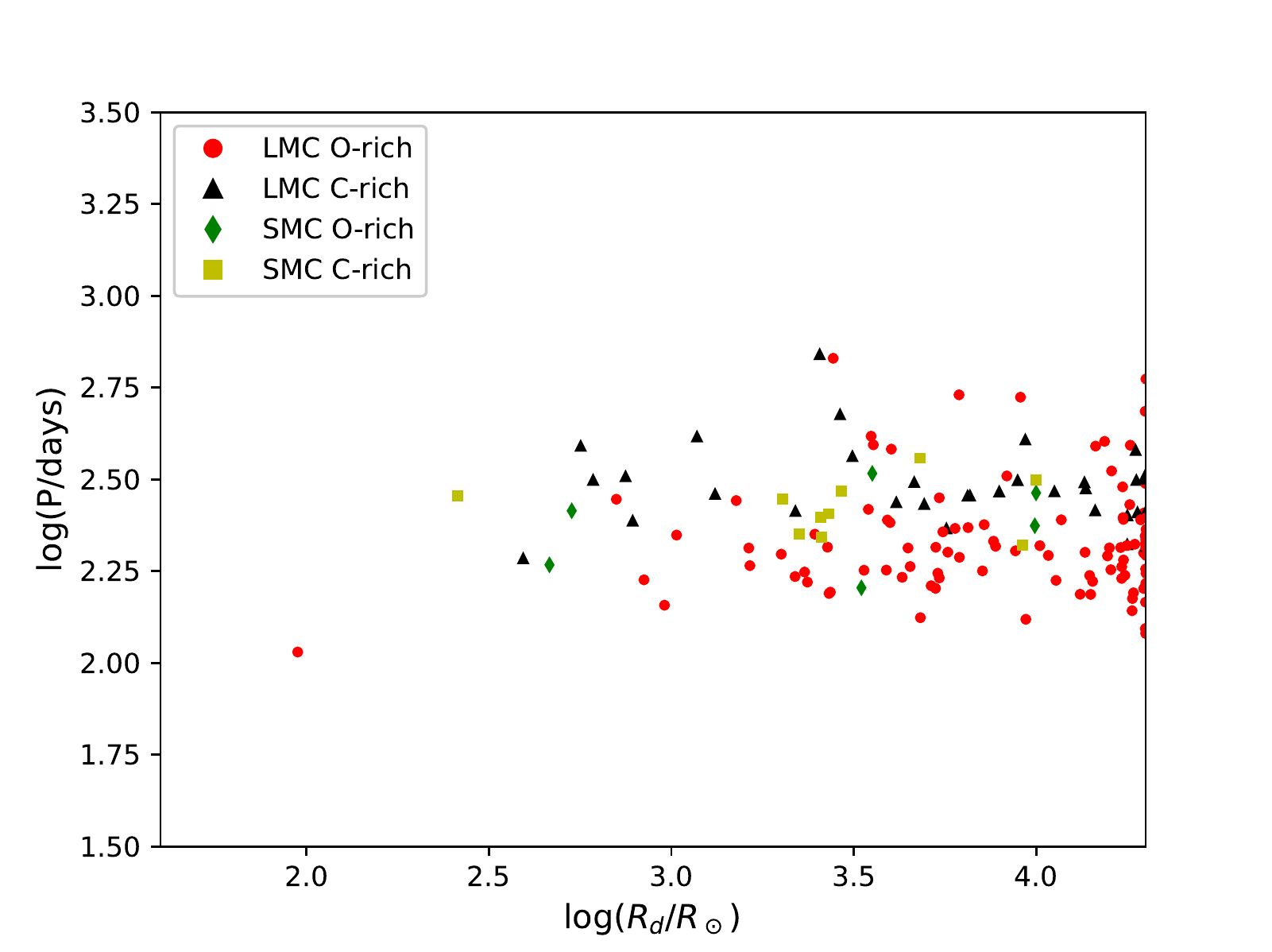}
    \caption{The period-radius relation for regular Miras, separated for the stellar component (left panel) and the dust component (right panel), based on the results of our SEDs fitting (see Table \ref{fit_data}). The dash line on the left panel is the "death line" adopted from \citet{2021MNRAS.500.1575T}.} 
    \label{PR}
\end{figure*}

\begin{figure*}
    \centering
    \begin{tabular}{ccc}
    \includegraphics[width=0.31\columnwidth]{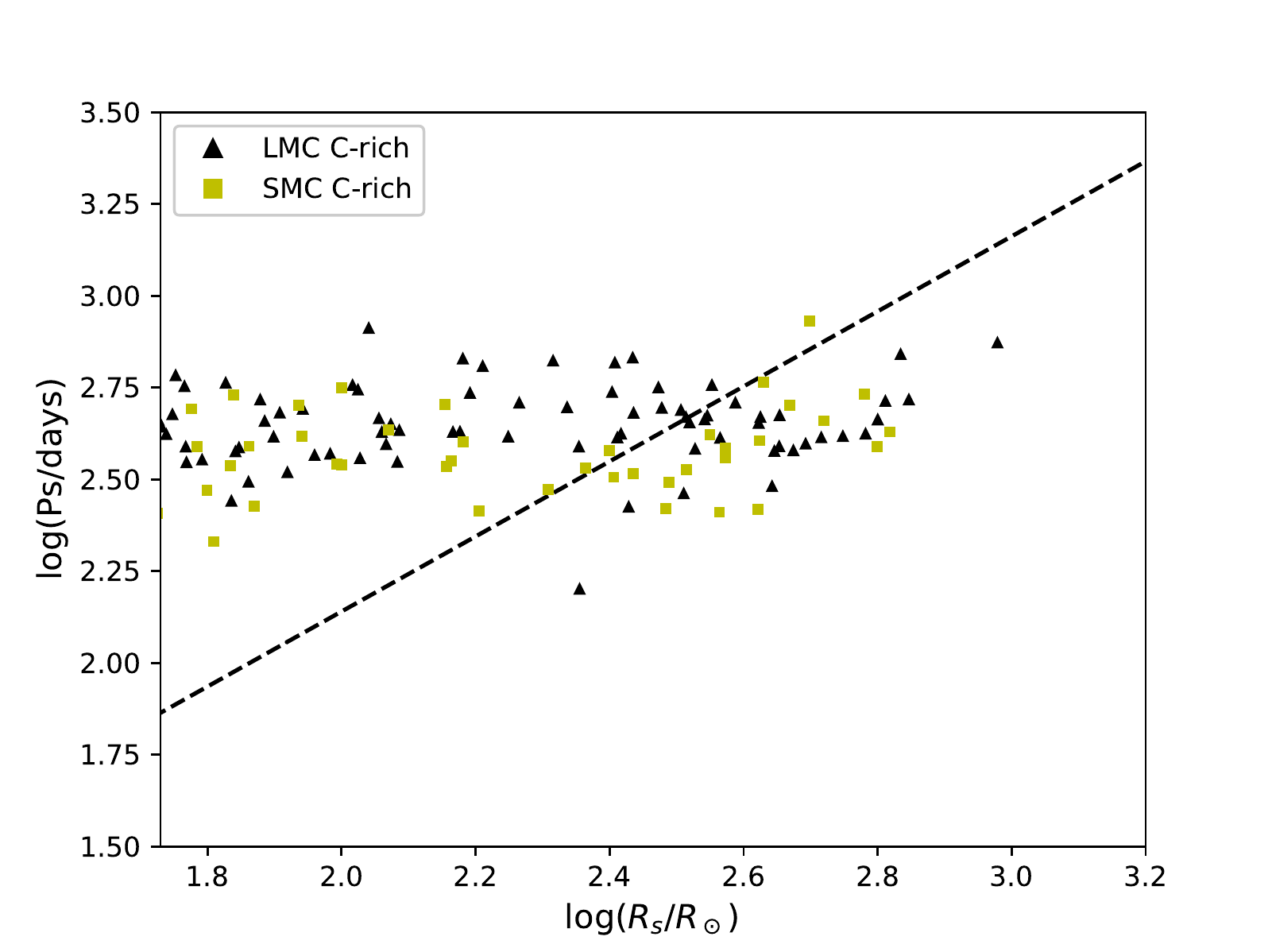} & \includegraphics[width=0.31\columnwidth]{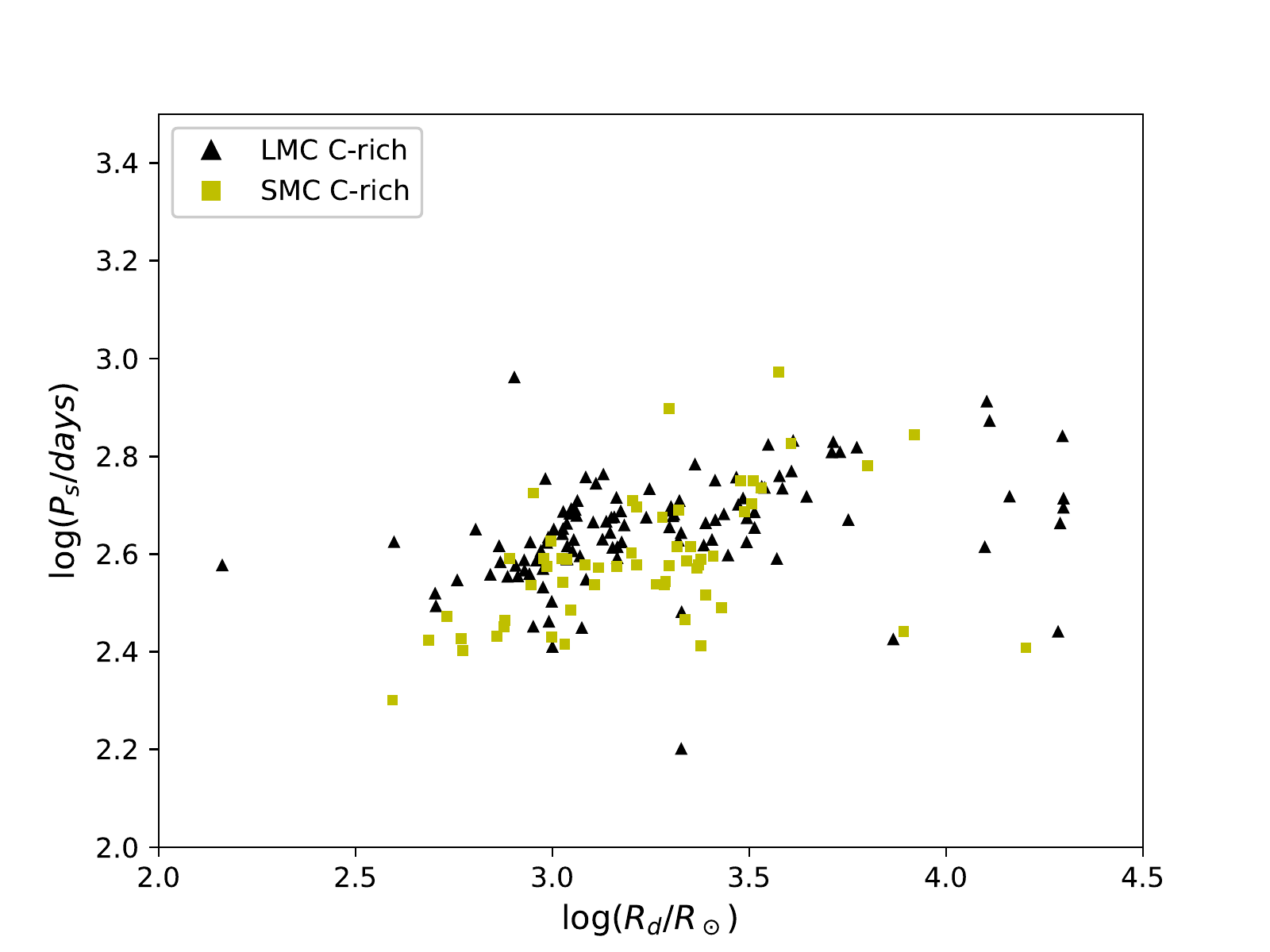} & \includegraphics[width=0.31\columnwidth]{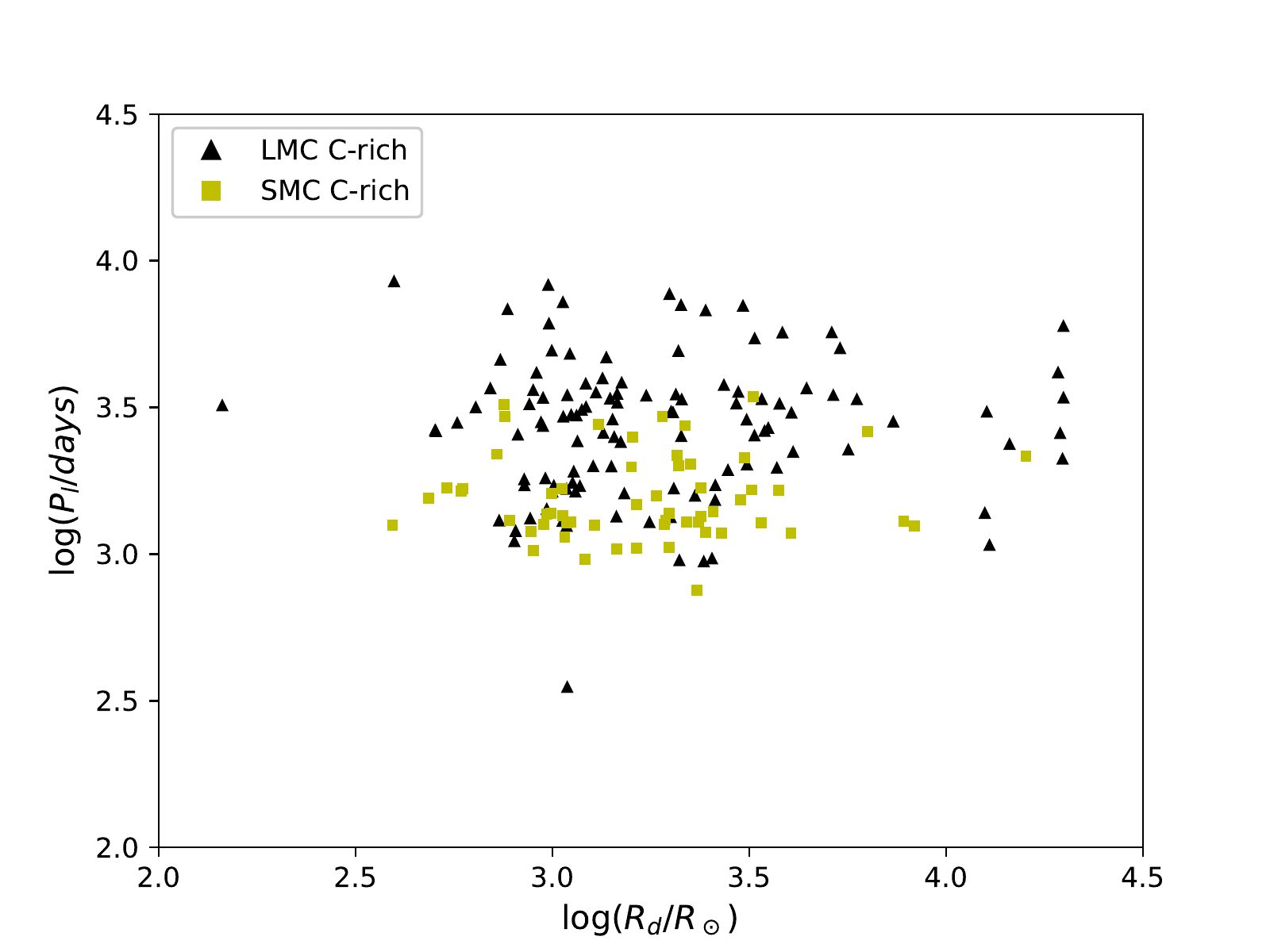} 
    \end{tabular}
    \caption{The period-radius relation for non-regular Miras, separated for the stellar component with $P_s$ (left panel), the dust component with $P_s$ (middle panel), and the dust component with $P_l$ (right panel), based on the results of our SEDs fitting (see Table \ref{fit_data}). The dash line on the left panel is the "death line" adopted from \citet{2021MNRAS.500.1575T}.}
    \label{PR_multi}
\end{figure*}

Some by-products obtained from the two-component black-body SED fitting are the radii of the stellar and dust components. In Table \ref{fit_data}, we summarized the fitted temperatures (in Kelvin) and radius (in Solar radius) for both components in our sample of regular Miras. The left panel of Figure \ref{PR} presents the corresponding period-radius (PR) relation for the regular Miras, at which the regular Miras followed a well-defined PR relation. In the left panel of Figure \ref{PR}, we included a "death line", as defined in \citet{2021MNRAS.500.1575T}, to represent the maximum pulsation periods for a given stellar radius based on a series of theoretical model calculations. The regular Miras are located below this line. In the right panel of Figure \ref{PR}, we presented a similar PR relation for the dust components, wherein no correlation was found between the pulsation periods of regular Miras and the radii of the dust components. In Figure \ref{PR_multi}, we present the PR relations for the C-rich non-regular Miras that is similar to Figure \ref{PR}, where the $P_s$ and $P_l$ are adopted from Table \ref{fitper_data}.  There seems to have some trends between $R_d$ and $P_s$, for $\log R_d$ between $2.5$ and $4.0$, in case of the non-regular Miras,\footnote{Whether such trends are real or not has to be investigated in details using a more realistic model, which is beyond the scope of this paper.} but in general, no obvious correlations were found between the radii and periods, neither for the stellar nor the dust component based on our SED fitting results.

 


\section{Summary}

In this work, we classified the Magellanic Clouds Miras into 694 regular Mira and 1321 non-regular Miras based on their OGLE-III $I$-band light curves through machine learning.  For regular Miras, we found that, at maximum light, their magnitude is relatively stable, while at minimum light, a large fraction exhibit periodic variations. Dispersions of the PL relation at maximum light is smaller for O-rich, while C-rich is larger for their corresponding mean light counterparts.

For the non-regular Miras, we simultaneously determined the periods corresponding to the short- and long-term variations and no clear correlation was seen between these periods and the apparent magnitudes even after removing the long-term variations in their light curves. Our results also suggest the (periodic) long-term variations are not directly associated with pulsation because after removing the long-term trends, the O-rich non-regular Miras was located close to the PL relation fitted from the regular Miras (Figure \ref{rmpl}).

Using the available multi-band photometry, we performed SED fitting of our sample of Miras with two black-body radiation functions for the stellar and dust components. Results based on our SED fitting, together with the evidence from the CMD and the $JHK$-band color-color diagram, showed that a large abundance of dust could be found in the non-regular Miras. In contrast, regular Miras did not exhibit evidence for the presence of dust. This suggested that a large fraction of dust found in the non-regular Miras could be responsible for their long-term variations shown in the light curves.


\acknowledgments

We thank the useful discussions and comments from an anonymous referee to improve the manuscript. We thank the funding from Ministry of Science and Technology (Taiwan) under the contract 109-2112-M-008-014-MY3.

This publication makes use of data products from the Two Micron All Sky Survey, which is a joint project of the University of Massachusetts and the Infrared Processing and Analysis Center/California Institute of Technology, funded by the National Aeronautics and Space Administration and the National Science Foundation.


\software{{\tt astropy} \citep{astropy2013,astropy2018}, {\tt FATs} \citep{nun15, 2017ascl.soft11017N}, {\tt Scikit-learn} \citep{JLMR}, {\tt DESK} \citep{2020JOSS....5.2554G}.}

\end{document}